\documentclass[pra,twocolumn,superscriptaddress,notitlepage,showpacs,showkeys]{revtex4-1}
\usepackage{graphicx,amsmath,amsfonts,amssymb,upgreek,txfonts,color}
\usepackage[colorlinks,linkcolor=blue,citecolor=blue,urlcolor=blue,breaklinks=true]{hyperref}

\begin{document}

\title{Strong quadrature squeezing and quantum amplification in a coupled Bose-Einstein condensate- optomechanical cavity via coherent modulation
}

\author{Ali Motazedifard} 
\email{motazedifard.ali@gmail.com}
\address{Quantum Optics Group, Department of Physics, Faculty of Science, University of Isfahan, Hezar Jerib, 81746-73441, Isfahan, Iran}

\author{A. Dalafi} 
\email{a\_dalafi@sbu.ac.ir}
\address{Laser and Plasma Research Institute, Shahid Beheshti University, Tehran 19839-69411, Iran}

\author{M. H. Naderi} 
\email{mhnaderi@sci.ui.ac.ir}
\address{Quantum Optics Group, Department of Physics, Faculty of Science, University of Isfahan, Hezar Jerib, 81746-73441, Isfahan, Iran}

\author{R. Roknizadeh}
\email{rokni@sci.ui.ac.ir}
\address{Quantum Optics Group, Department of Physics, Faculty of Science, University of Isfahan, Hezar Jerib, 81746-73441, Isfahan, Iran}


\date{\today}
\begin{abstract}
We consider an optomechanical cavity with a movable end-mirror as a quantum mechanical oscillator (MO) containing an interacting cigar-shaped Bose-Eisenstein condensate (BEC). It is assumed that both the MO and the BEC interact with the radiation pressure of the cavity field in the red-detuned and weak coupling regimes while the two-body atomic collisions frequency of the BEC and the mechanical spring coefficient of the MO are coherently modulated. By analyzing the scattering matrix we find that a frequency-dependent squeezing is induced to the three subsystems only due to the coherent modulations. In the largely different cooperativities regime together with strong modulations, the mechanical mode of the MO and the Bogoliubov mode of the BEC exhibit quadrature squeezing which can surpass the so-called 3dB limit (up to 75 dB) with high robustness to the thermal noises. Surprisingly, in this regime by controlling the system and modulation parameters, a very high degree of squeezing (up to 16 dB) together with high purity of quantum state for the output cavity field is achievable. Furthermore, one can attain simultaneous strong quantum amplification, added-noise suppression, and controllable gain-bandwidth for the complementary quadratures of squeezed ones in the subsystems.

\end{abstract}


\maketitle

\section{Introduction}
Over the past decade, there has been a significant theoretical and experimental progress in the field of quantum optomechanics which deals with linear or quadratic coupling of the electromagnetic radiation pressure to a mechanical oscillator (MO) \cite{Aspelmeyer,Meystreoptomechanics,chenoptomechanics,milburnoptomechanics}. Quantum optomechanical systems (OMSs) have had a remarkable contribution to the emergence of quantum behaviors on macroscopic scales and have made a significant contribution in probing fundamentals of physics. OMSs have many practical applications such as displacement and force sensing \cite{xsensing1,CQNCPRL,CQNCmeystre,aliNJP,complexCQNC}, ground-state cooling of the MO \cite{ground state cooling, Sideband cooling,Laser cooling,Sideband cooling,Chan}, generation of entanglement \cite{Palomaki2,Paternostro,genes-entangelment}, synchronization of MOs \cite{Mari1,MianZhang,Bagheri,grebogi,foroud} and generation of nonclassical states of the mechanical and optical modes \cite{Borkje,Hammerer,foroudnonlinearcoherent,unconditionalstate,adde-coherentstate}. Very recently, OMSs have been proposed for the realization of the \textit{parametric}-dynamical Casimir effect (\textit{P}-DCE) \cite{DCEQWell,aliDCE1,aliDCE2,aliDCE3} and the quantum simulation of the curved space-times \cite{foroudCurvedspacetime}.


Among various applications of OMSs, quantum amplification \cite{Refrevise2,Refrevise3,Refrevise4,Refrevise5} and generation of squeezing \cite{Refrevise6,Refrevise7,Kippenbergsqueezing} have been very attractive fields of research in the realm of quantum optics over the recent years. In addition to the standard ponderomotive squeezing \cite{Aspelmeyer,ponderomotive,vitaliponderomotive} and more recently proposed method of dissipative-squeezing \cite{Clerkdissipativeoptomechanics} and also methods of Refs.~\cite{Refrevise2,Refrevise3,Refrevise4,Refrevise5,Refrevise6,Refrevise7}, there is a well-known method based on the modulation technique to achieve the quadrature squeezing in OMSs \cite{aliDCE1,aliDCE2,gentlymodulating,giovanettimodulation,lawmodulation,schmidtmodulation,milburnmodulation,pontinmodulation} which at best, can lead to $50\%$ noise reduction below the zero-point level (the so-called $ 3 $ dB limit). Nevertheless, noise reduction much below the standard quantum limit (SQL) is also achievable by parametrically modulating the spring constant of
the MO in a bare optomechanical cavity \cite{optomechanicswithtwophonondriving}. As has been shown in Ref. \cite{optomechanicswithtwophonondriving} through the mentioned modulation method the system can also act as an optical quantum amplifier. Besides, there are some methods based on the feedback process \cite{clerkfeedback}, quantum measurement \cite{Harris,Bowen}, and quadratic optomechanical coupling (QOC) \cite{Nunnenkamp,kumar,asjad,dalafiQOC} in order to surpass and beat the $ 3$dB limit. Also, very recently, it has been shown \cite{twofoldsqueezing} that the modulation of the amplitude of the driving laser in an optomechanical cavity containing an optical parametric amplifier can generate two-fold mechanical squeezing beyond the SQL.


On the other hand, hybrid OMSs assisted with trapped Bose-Einstein condensates (BECs) \cite{MeystreBEC,BrennBECexp,RitterBECexp} or ultra cold atoms integrated in an atom chip \cite{Refrevise1}  in which the fluctuations in the collective excitations of the atoms, i.e., the so-called Bogoliubov mode, plays the role of an effective mechanical mode while the nonlinear atom-atom interaction plays the role of an atomic amplifier \cite{dalafi1,dalafi3,dalafi4}, have attracted considerable attention to a number of studies such as normal mode splitting (NMS) \cite{BhattacherjeeNMS}, noise reduction \cite{Bhattacherjeenoisereduction,dalafi5}, and controlling bistability, cooling, and bipartite entanglement \cite{dalafi2,dalafi6,dalafi7,dalafi8}. Moreover, very recently, it has been shown \cite{dalafiQOC} that through the effect of the QOC in the OMS assisted with the BEC, one can achieve a considerable degree of squeezing (up to 10 dB) in the mechanical mode and a strong stationary entanglement between the mechanical and atomic modes.


Motivated by the above-mentioned studies as well as inspired by the proposed scheme of Ref.~\cite{optomechanicswithtwophonondriving}, in the present paper we are going to investigate a hybrid optomechanical system as a quantum linear amplifier/squeezer with an extra degree of freedom (the Bogoliubov mode of the BEC) in addition to the mechanical mode of the MO in which both the mechanical spring coefficient of the MO and the two-body atomic collisions frequency of the BEC are coherently modulated. The advantage of the present hybrid system in comparison to the bare one presented in Ref.~\cite{optomechanicswithtwophonondriving} is that the presence of the extra Bogoliubov mode of the BEC which have much more controllability with respect to the mechanical mode of the MO, can lead to a considerable enhancement of the squeezing and amplification in both the mechanical and the optical modes.

On the other hand, we have shown in a very recent work \cite{aliDCE3} that the parametric modulations of the mechanical and atomic modes in such a hybrid system can lead to the realization of the parametric DCE of photons and phonons. However, in the present paper we investigate some other features of the system, including controllable realization of strong quadrature squeezing together with complementary-quadrature amplification, added-noise suppression, and controllable increasing of the gain-bandwidth. In other words, we study the system as a controllable optomechanical squeezer/amplifier.

Here, by determining the quadratures spectra with use of the scattering-matrix method we find that in the so-called regime of largely different cooperativities high degrees of quadrature squeezing for both the output cavity field (up to 16 dB) and phonon modes (up to 75 dB ) can be achieved through the modulation of the atomic collisions frequency and the mechanical spring coefficient. Such high degrees of quadratures squeezing are originated from a frequency-dependent squeezing effect induced to the cavity field, the MO, and the BEC. Besides, It is shown that the complementary quadratures of squeezed ones, can be strongly amplified in the large driving regime where the added noises can be strongly suppressed simultaneously. In this situation the system behaves as a noiseless phase-sensitive amplifier. Furthermore, their gain-bandwidth can be externally controlled by adjusting the amplitudes of modulations and atomic collisions frequency. 

The paper is organized as follows. In Sec.~\ref{sec.system}, we describe the physical system and its effective Hamiltonian.
In Sec.~\ref{sec.dynamics}, by solving the quantum Langevin equations in frequency space we find the self-energies and induced-parametric squeezing coefficients. In Sec.~\ref{sec.Scatt} we investigate the quadratures amplification spectra and the gain-bandwith using the scattering matrix method. In Sec.~\ref{sec.squeezing} by calculating the quadratures squeezing spectra we discuss how one can achieve high degrees of quadrature squeezing for both photon and phonon modes. Finally, we summarize our conclusions in Sec.~\ref{sec.summary}

\section{The System Hamiltonian \label{sec.system}}
\begin{figure}
	\centering
	\includegraphics[width=8.7cm]{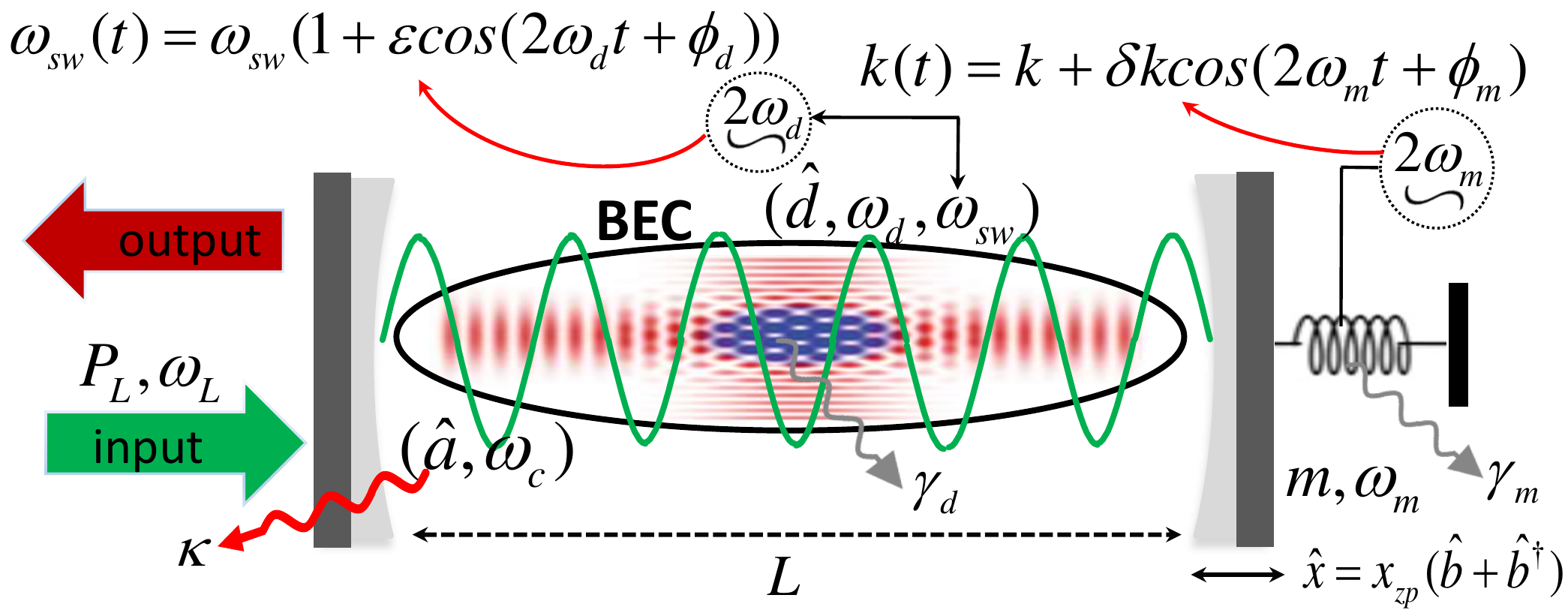} 
	\caption{(Color online) Schematic of an atomic BEC-hybridized red-detuned driven optomechanical cavity with a movable end mirror whose mechanical spring coefficient, $ k(t) $, and the atomic collisions frequency, $ \omega_{sw}(t) $, are modulated parametrically. $ \kappa $, $ \gamma_m $ and $ \gamma_d $ are the cavity, the mechanical and Bogoliubov damping rates, respectively.}
	\label{fig1}
\end{figure}

As shown in Fig.~(\ref{fig1}), we consider a red-detuned driven \textit{hybrid} optomechanical cavity with length $ L $ whose end mirror with effective mass $ m $ and damping rate $ \gamma_m $ is free to oscillate at mechanical frequency  $ \omega_m $. The cavity contains a cigar-shaped BEC of $ N $ ultracold two-level atoms with mass $ m_a $ and transition frequency $ \omega_a $. The cavity is driven at rate $E_L= \sqrt{\kappa P_L / \hbar \omega_L}$ through the fixed mirror by a laser with frequency $\omega_{L}$ and wavenumber $k_{0}=\omega_{L}/c$ ($P_{L}$ is the laser power and $\kappa$ is the cavity decay rate). 

We assume that the MO is parametrically driven by modulating its spring coefficient at twice its natural frequency \cite{optomechanicswithtwophonondriving,aliDCE3}, i.e., $k(t)= k+\delta k \cos(2\omega_{m} t +\phi_m) $ ($ \phi_m $ being the phase of external modulation). Under these assumptions, the total Hamiltonian of the system in the frame rotating at laser frequency is given by
\begin{eqnarray} \label{H1tot}
&& \hat H= \hat H_0 + \hat H_{om} + \hat H_m(t) + i \hbar E_L (\hat a-\hat a^\dag ) + \hat H_{BEC}(t).
\end{eqnarray}
The first term in Eq.~(\ref{H1tot}) describes the free Hamiltonian of the cavity field and the MO, the second term describes the optomechanical interaction, and the third term corresponds to the mechanical parametric drive whose explicit forms are, respectively, given by
\begin{subequations}
	\begin{eqnarray}
	&& \hat H_{0}=\hbar \Delta_c \hat a^\dag \hat a + \hbar \omega_m \hat b^\dag \hat b , \label{H_0}\\
	&& \hat H_{om}= -\hbar g_0 \hat a^\dag \hat a (\hat b + \hat b^\dag), \label{H_om} \\
	&& H_m(t)= \frac{i \hbar}{2} (\lambda_m \hat b^{\dag 2}  e^{-2i\omega_m t}- \lambda_m^\ast \hat b^2 e^{2i\omega_m t}), \label{H_mod_t}
	\end{eqnarray}
	\end{subequations}
where $ \hat a $ ($ \hat b $) is the annihilation operator of the cavity (mechanical) mode, $ \Delta_c= \omega_c-\omega_L $ is the detuning of the cavity mode from the driving laser frequency, and $ g_0=x_{zp} \omega_c /L $ stands for the single-photon optomechanical coupling with $ x_{zp}= \sqrt{\hbar / 2m\omega_m} $ being the mechanical zero-point position fluctuation. 
Besides, $ \lambda_m= \vert \lambda_m \vert e^{i\phi_m} $ characterizes the strength and phase of the mechanical parametric driving with $ \vert \lambda_m \vert= \delta k x_{zp}^2 / 2\hbar $, which can be taken real by fixing the phase of modulation $ \phi_m $. Note that the Hamiltonian of Eq.~(\ref{H_mod_t}) describes the mechanical phonon analog of the degenerate parametric amplification (DPA) where the vibrational fluctuation of the MO plays the role of the signal mode in the DPA. In Ref.~\cite{aliDCE3}, it has been shown that such a phonon analog of DPA leads to the amplification of quantum vacuum fluctuations of the MO, i.e., the DCE of mechanical phonons.

The last term in Eq.~(\ref{H1tot}) describes the many-body Hamiltonian of the BEC in the framework of second quantization \cite{dalafi2,dalafi6,dalafi7,dalafi8}. In the dispersive regime of atom-field interaction where the laser pump is far-detuned from the atomic resonance ($ \Delta_a=\omega_a - \omega_L \gg \Gamma_a $ where $ \Gamma_a$ is the atomic linewidth)  the Hamiltonian of the BEC can be written as \cite{Ritsch}
\begin{eqnarray} \label{H1_BEC}
&& \hat  H_{BEC}=\int_{-L/2}^{L/2}dx \hat \psi^\dag(x) \Big[\frac{-\hbar^2}{2m_a} \frac{d^2}{dx^2},\nonumber\\ 
&& + \hbar U_0 \cos^2(k_0 x) \hat a^\dag \hat a+\frac{1}{2}U_{s} \hat \psi^{\dag}(x)  \hat \psi^{\dag}(x)\Big] \hat \psi(x),
\end{eqnarray}
where $ \hat \psi(x) $ is the annihilation operator of the atomic field, $ U_0=-g_a^2/\Delta_a $ is the optical lattice barrier height per photon which represents the atomic backaction on the field in the dispersive regime, $ g_a $ is the vacuum Rabi frequency or atom-field coupling constant, $ U_s=4\pi \hbar^2 a_s/m_a $, and $ a_s $ is the two-body \textit{s}-wave scattering length \cite{Ritsch,Domokos2}. In the weakly interacting regime \cite{dalafi1}, and under the Bogoliubov approximation \cite{Nagy} the atomic field operator can be expanded as the following single-mode quantum field
\begin{eqnarray} \label{si1}
&& \hat \psi(x)= \sqrt{N/L}+ \sqrt{2/L}\cos(2k_0x) \hat d ,
\end{eqnarray}
where the so-called Bogoliubov mode $ \hat d $ corresponds to the quantum fluctuations of the atomic field around the classical condensate mode $ \sqrt{N/L} $.

On the other hand, as has been shown in Ref.~\cite{JaskulaBECmodulationexp}, the time modulation of the \textit{s}-wave scattering frequency of atom-atom interaction, $ \omega_{sw} $, can be realized by the time modulation of the scattering length via the modulation of the electromagnetic trap, or the modulation of the density of the BEC by changing the trap stiffness via the intensity modulation of the pump laser. Now, considering the time modulation of the \textit{s}-wave scattering frequency as $ \omega_{sw}(t)=\omega_{sw}\Big(1+\epsilon\cos(2\omega_{d}t+\phi_{d})\Big) $  where $ \epsilon $ and $ \phi_d $ refer, respectively, to the amplitude and the phase of modulation, $ \omega_{d}=4\omega_{R}+\omega_{sw} $ with $ \omega_{R}=\hbar k_{0}^{2}/2m_a $ being the recoil frequency and $ \omega_{sw}=8\pi\hbar N a_s/(m_a L w^2) $ ($ w $ is the waist radius of the optical mode), the Hamiltonian of the modulated BEC within the rotating wave approximation (RWA) can be written as \cite{aliDCE3}
\begin{subequations}
	\begin{eqnarray}
	&&  \hat H_{BEC}=\hbar\delta_0 \hat a^\dag \hat a  +  \hbar \omega_d \hat d^\dag \hat d + \hbar G_0\hat a^\dag \hat a (\hat d  + \hat d^\dag) +  \! \hat H_{sw}(t) , \label{H_BEC} \label{Hbec}\\
	&& \hat H_{sw}(t)=\frac{i \hbar}{2} (\lambda_d \hat d^{\dag 2}  e^{-2i\omega_d t}- \lambda_d^\ast \hat d^2 e^{2i\omega_d t}), \label{H_sw}
	\end{eqnarray}
	\end{subequations}
where  $ \delta_0=NU_0/2 $ and $ \lambda_{d}=-i\epsilon\omega_{sw} e^{-i\phi_{d}}/4 $ can be taken real by fixing the phase $ \phi_{d} $. Furthermore, $ G_0=\sqrt{2N}U_0/4 $ is the strength of an optomechanical-like coupling between the Bogoliubov mode of the BEC and the intracavity field . The first term of the Hamiltonian of Eq.~(\ref{Hbec}) leads to a shift of the cavity detuning,i.e., $ \Delta_c \to \Delta_0=\Delta_c + NU_0/2 $ which can be interpreted as an effective Stark-shifted detuning.

\section{Dynamics of the system \label{sec.dynamics}}
Here, we investigate the dynamics of the system under the conditions that the frequency of the Bogoliubov mode of the BEC, $ \omega_d $, can be matched to the mechanical frequency ($ \omega_d \approx \omega_m $) via the atomic collisions frequency $ \omega_{sw} $ or through the recoil frequency $ \omega_R $ via the driving laser frequency. Therefore, we restrict ourselves to the red-detuned regime of cavity optomechanics where $ \Delta_0 \approx \omega_m \approx \omega_d $. In this regime and within the RWA where the BEC-cavity mode as well as the MO-cavity mode couplings are analogous to the beam-splitter interaction, the linearized QLEs describing the dynamics of the quantum fluctuations are given by ~\cite{aliDCE3}
\begin{subequations} \label{QLEBS}
	\begin{eqnarray}
	&& \delta \dot { \hat a} = -\frac{\kappa}{2} \delta \hat a + i g  \delta \hat b - i G \delta \hat d + \sqrt{\kappa} \hat a_{in} , \\
	&& \delta \dot {\hat b}  = -\frac{\gamma_m}{2} \delta \hat b + i g \delta \hat a + \lambda_m \delta \hat b^\dag +  \sqrt{\gamma_m} \hat b_{in}, \\
	&& \delta \dot {\hat d}  = - \dfrac{\gamma_d}{2} \delta \hat d  -i G  \delta \hat a + \lambda_d \delta \hat d^\dag  +  \sqrt{\gamma_d} \hat d_{in},
	\end{eqnarray}
\end{subequations}
where $ g=g_0 \bar a $ and $ G = G_0 \bar a $ are, respectively, the enhanced-optomechanical coupling strengths of the moving mirror and the Bogoliubov mode of the BEC to the intracavity field with $\bar a $ being the optical mean-field value. Furthermore, $\gamma_{m}$ and $\gamma_{d}$  denote the dissipation rates of the mechanical and the Bogoliubov modes, respectively.

According to Eqs.(\ref{QLEBS}a)-(\ref{QLEBS}c), three noise operators affect the system: the optical input vacuum noise, $a_{in}$, and the thermal Brownian noise operators $b_{in}$ and $d_{in}$ acting, respectively, on the MO and the Bogoliubov mode of the BEC. These noises which are assumed to be uncorrelated for the different modes of both the matter and light fields, satisfy the Markovian correlation functions \cite{gard,giov} $\langle\hat a_{in}(t)\hat a_{in}^{\dagger}(t^{\prime})\rangle=(\bar n_{a}^T+1)\delta(t-t^{\prime})$, $ \langle\hat a_{in}^{\dagger}(t)\hat a_{in}(t^{\prime})\rangle=\bar n_{a}^T\delta(t-t^{\prime}) $,  $ \langle\hat b_{in}(t)\hat b_{in}^{\dagger}(t^{\prime})\rangle=(\bar n_{m}^T+1)\delta(t-t^{\prime}) $, $ \langle\hat b_{in}^{\dagger}(t)\hat b_{in}(t^{\prime})\rangle=\bar n_{m}^T\delta(t-t^{\prime}) $,
$ \langle\hat d_{in}(t)\hat d_{in}^{\dagger}(t^{\prime})\rangle=(\bar n_{d}^T+1)\delta(t-t^{\prime}) $, and $ \langle\hat d_{in}^{\dagger}(t)\hat d_{in}(t^{\prime})\rangle=\bar n_{d}^T\delta(t-t^{\prime}) $ where $ \bar n_j^T=[\exp(\hbar \omega_j/k_B T_j)-1]^{-1} $ with $j=a,m,d$ are the thermal excitations of the optical, mechanical and Bogoliubov modes at temperature $T_j$.

Now, by defining the fluctuation quadratures $ \delta \hat X_o=(\delta\hat o+\delta\hat o^{\dagger})/\sqrt{2} $ and $ \delta\hat P_o=(\delta\hat o-\delta\hat o^{\dagger})/\sqrt{2}i $ ($ o=a,b,d $) we can express the linearized Eqs.~(\ref{QLEBS}a)-(\ref{QLEBS}c) in the compact matrix form as
\begin{eqnarray} \label{udot}
&&\delta \dot {\hat u}(t)= A ~\delta \hat u(t) + \hat u_{in}(t),
\end{eqnarray}
where the vector of continuous-variable fluctuation operators and the corresponding vector of noises are given by $ \delta\hat u=(\delta \hat X_a,\delta \hat P_a,\delta \hat X_b, \delta \hat P_b,\delta \hat X_d,\delta \hat P_d)^T $ and $ \hat u_{in}=(\sqrt{\kappa}\hat X_a^{in},\sqrt{\kappa}\hat P_a^{in},\sqrt{\gamma_m}\hat X_b^{in},\sqrt{\gamma_m}\hat P_b^{in}, \sqrt{\gamma_d} \hat X_d^{in},\sqrt{\gamma_d} \hat P_d^{in} )^T $, respectively. Here, $ \hat X_o^{in}=(\hat o_{in}+\hat o_{in}^{\dagger})/\sqrt{2} $ and $ \hat P_o^{in}=(\hat o_{in}-\hat o_{in}^{\dagger})/\sqrt{2}i $ ($ o=a,b,d $).
Furthermore, the time-independent drift matrix $ A $ is given by 
\begin{eqnarray} \label{A}
&&\!\!\!\!\!\!\!\!\!\!\!\!  A\!=\! \left( \begin{matrix}
{-\frac{\kappa}{2}} & {0} & {0} & {-g} & {0} & {G}  \\
{0} & {-\frac{\kappa}{2}} & {g} & {0} & {-G} & {0}  \\
{0} & {-g} & {\!\lambda_m\!-\!\frac{\gamma_m}{2}} & {0} & {0} & {0}  \\
{g} & {0} & {0} & {-(\lambda_m\!+\!\frac{\gamma_m}{2})} & {0} & {0}  \\
{0} & {G} & {0} & {0} & {\lambda_d -\frac{\gamma_d}{2}} & {0}  \\
{-G} & {0} & {0} & {0} & {0} & {-(\lambda_d\!+\!\frac{\gamma_d}{2} )}  \\
\end{matrix} \right).
\end{eqnarray}


In the following sections we will show how modulation of the atomic collisions frequency or the mechanical spring coefficient leads to the strong quadrature squeezing of the mechanical and Bogoliubov modes. As has been shown in Ref.~\cite{aliDCE3}, based on the Routh-Hurwitz criterion for the optomechanical stability condition, the modulation parameters $ \lambda_{m} $ and $ \lambda_{d} $ should satisfy the condition
\begin{eqnarray} \label{stability}
&&\!\!\!\!\!\!\! \lambda_{m(d)} \le \lambda_{m(d)}^{max}=\gamma_{m(d)} \! + \! \Gamma_{op}^{m(d)}= \! \frac{\gamma_{m(d)}}{2} \left[1+ \mathcal{C}_{m(d)}\right],
\end{eqnarray}
where the maximum optomechanically induced damping rate is $ \Gamma_{op}^{m(d)}=-2{\rm Im} \Sigma_{b(d)}(0) $ in which the self-energies $\Sigma_{b(d)}$ have been given by Eqs.~(26) of Ref.~\cite{aliDCE3}. Furthermore, $ \mathcal{C}_m $($ \mathcal{C}_d $) is the collective optomechanical cooperativity which is given by 
\begin{eqnarray}
&&  \mathcal{C}_{m(d)}=  \mathcal{C}_{0(1)} \frac{1+ \mathcal{C}_{1(0)} - \xi_{d(m)}^2}{(1+ \mathcal{C}_{1(0)}-\xi_{d(m)}^2)^2- \xi_{d(m)}^2  \mathcal{C}_{1(0)}^2} ~,
\end{eqnarray}
where $\mathcal{C}_0=4g^2/\kappa \gamma_m$ and $\mathcal{C}_1= 4G^2/\kappa \gamma_d $ are, respectively, the optomechanical and opto-atomic cooperativities, and $ \xi_{d(m)}=2\lambda_{d(m)}/\gamma_{d(m)}$ plays the role of an effective dimensionless amplitude of modulation. 

Before calculating the scattering matrix in the next section it is worthwhile, here, to consider the solutions of the linearized QLEs [Eqs.~(\ref{QLEBS}a)-(\ref{QLEBS}c)] in frequency space which are given by \cite{aliDCE3}
\begin{subequations}
	\begin{eqnarray}
	&& \!\! -i\omega \delta \hat a  =\! -[\kappa/2 \! +\! i\Sigma_a(\omega)] \delta \hat a \! + \!  \tilde \lambda_a \delta \hat a^\dagger \! + \!\!\! \sqrt{\kappa} \hat A_{in}(\omega) , \label{a_omega} \\
	&&\!\! -i\omega \delta \hat b=\! -[\gamma_m/2 \!- \! i\Sigma_b(\omega)] \delta \hat b \! + \! \tilde \lambda_b \delta \hat b^\dagger \! + \!\!\! \sqrt{\gamma_m} \hat B_{in}(\omega), \label{b_omega} \\
	&& \!\! -i\omega \delta \hat d= \! -[\gamma_d/2 \! + \! i \Sigma_d(\omega)] \delta \hat d \! + \! \tilde \lambda_d \delta\hat d^\dagger\! + \!\!\! \sqrt{\gamma_d} \hat D_{in}(\omega). \label{d_omega} 
	\end{eqnarray}
\end{subequations}
Here, $\Sigma_o(\omega)$ (o=a,b,d) is the self-energy (given by Eq.~(26) in Ref.~\cite{aliDCE3}), $\hat{A}_{in}, \hat{B}_{in}$, and $\hat{D}_{in}$ denote the generalized noise operators (given by Eq.~(29) in Ref.~\cite{aliDCE3}), and the induced frequency-dependent parametric amplification coefficients $\tilde\lambda_o(\omega) $ which are analogous to the gain factors in the conventional DPA are given by \cite{aliDCE3}
\begin{subequations}
	\begin{eqnarray}
	&&\!\!\!\!\!\!\!\!\! \tilde \lambda_a(\omega)=\frac{ g^2  \lambda_m}{(\gamma_m/2-i \omega)^2-\vert \lambda_m \vert^2} + \frac{G^2 \lambda_d}{(\gamma_d/2-i \omega)^2-\vert \lambda_d \vert^2} , \label{lambda_a_omega} \\
	&&\!\!\!\!\!\!\!\!\! \tilde \lambda_b(\omega)= \lambda_m + g^2 \frac{\bar \Lambda_s(\omega)}{[\kappa/2-i(\omega-\Sigma_s(\omega))]^2- \bar \Lambda_s^2(\omega)} ~, \label{lambda_b_omega} \\
	&&\!\!\!\!\!\!\!\!\! \tilde \lambda_d(\omega)= \lambda_{d} + G^2 \frac{\Lambda_m(\omega)}{[\kappa/2 - i(\omega-\Sigma_m(\omega))]^2 - \Lambda_m^2(\omega)} ~,  \label{lambda_d_omega}
	\end{eqnarray}
\end{subequations}
where the frequency-dependent functions $ \bar \Lambda_s $, $ \Lambda_m $, $\Sigma_m$ and $ \Sigma_s $ are given by Eq.~(27) in Ref.~\cite{aliDCE3}.
As is evident, by turning on the modulations (i.e., $\lambda_{m,d} \neq 0$), the mechanics and the atomic collisions do indeed mediate a parametric-amplifier-like effective squeezing interaction for the cavity mode which is also frequency-dependent (see Eq. (\ref{lambda_a_omega})), unlike the case of the conventional DPA. Furthermore, the time modulation of the atomic collisions (mechanical mode) induces indirectly the frequency-dependent squeezing through the second term in Eq.~(\ref{lambda_b_omega}) (Eq.~(\ref{lambda_d_omega})) to the mechanical (Bogoliubov) mode. As we will show in the next sections, such an induced-frequency-dependent squeezing is responsible for the strong quadrature squeezing in the subsystems. 

\section{quadratures amplification spectra \label{sec.Scatt}}
Using the theory of linear quantum amplifiers \cite{Refrevise8,IntQN}, in this and the following sections, we investigate the key role of the coherent modulation processes in the proposed hybrid optomechanical system as an amplifier/squeezer. For this purpose, it is more convenient to work in the frequency domain. Writing Eq.~(\ref{udot}) in the Fourier space gives
\begin{eqnarray} \label{u_omega}
&& \delta\hat u(\omega)=\boldsymbol{  { \chi }}(\omega) \sqrt{ { \Gamma}} \hat u_{in}(\omega) ,
\end{eqnarray}		
where $ \sqrt{ { \Gamma}} =  \rm {Diag}( \sqrt{\kappa},\sqrt{\kappa},\sqrt{\gamma_m},\sqrt{\gamma_m},\sqrt{\gamma_d},\sqrt{\gamma_d} ) $ and  the susceptibility matrix is given by
\begin{eqnarray} \label{susceptibility tensor}
&&\!\!\!\!\!\!\!\!\!\!\!\!\!  \boldsymbol{ { \chi }}(\omega) \! = \!  \left( \begin{matrix}
{\chi_{11}(\omega)} & {0} & {0} & {\chi_{14}(\omega)} & {0} & {\chi_{16}(\omega)}  \\
{0} & {\chi_{22}(\omega)} & {\chi_{23}(\omega)} & {0} & {\chi_{25}(\omega)} & {0}  \\
{0} & {\chi_{32}(\omega)} & {\chi_{33}(\omega)} & {0} & {\chi_{35}(\omega)} & {0}  \\
{\chi_{41}(\omega)} & {0} & {0} & {\chi_{44}(\omega)} & {0} & {\chi_{46}(\omega)}  \\
{0} & {\chi_{52}(\omega)} & {\chi_{53}(\omega)} & {0} & {\chi_{55}(\omega)} & {0}  \\
{\chi_{61}(\omega)} & {0} & {0} & {\chi_{64}(\omega)} & {0} & {\chi_{66}(\omega)}  \\
\end{matrix} \right)  \! ,
\end{eqnarray} 
where the matrix elements $ \chi_{ij}(\omega) $ have been given in the Appendix. In the absence of the BEC ($G_0=0, \lambda_{d}=0, \gamma_{d}=0$, and $\omega_{sw}=0$) $\chi(\omega)$ reduces to the susceptibility matrix given in Ref.~\cite{optomechanicswithtwophonondriving}.
Using input-output theory for the field operators, $ \hat o_{out}= \hat o_{in} + \sqrt{\gamma_o} \hat o  $ ($o=a,m,d$ with $\gamma_{a}=\kappa$), and going over to the quadrature basis, one can thus write
\begin{eqnarray} \label{u_out_omega}
&& \hat u_{out}[\omega]= s[\omega] \hat u_{in}[\omega],  
\end{eqnarray}
with the scattering matrix
\begin{eqnarray} \label{s}
&& s[\omega]= \hat I - \sqrt{  { \Gamma} }  \boldsymbol{ { \chi}}(\omega)  \sqrt{  { \Gamma} },
\end{eqnarray}
which can be simplified as 
\begin{eqnarray} \label{scattering}
	&& s[\omega]  = \left( \begin{matrix}
	{s_{11}} & {0} & {0} &  {s_{14} } & {0} & { s_{16}} \\
	{0} & {s_{22} }&{ s_{23}} & {0} & {s_{25}} & {0}  \\
	{0} & {s_{32}} & {s_{33}} & {0} & {s_{35}} & {0} \\
	{s_{41}} & {0} & {0} & {s_{44} }& {0} & {s_{46}} \\
	{0} & s_{52} & s_{53} & {0} & s_{55} & {0} \\
	s_{61}& {0} & {0} & s_{64} & {0} & s_{66} \\
	\end{matrix} \right),
\end{eqnarray} 
where $s_{11}(\omega)= 1 - \kappa \chi_{11}(\omega) $, $ s_{14}(\omega)=\sqrt{\kappa \gamma_m} \chi_{14}(\omega) $, 
$ s_{16}(\omega)= \sqrt{\kappa \gamma_d}\chi_{16}(\omega) $, $ s_{22}(\omega)={1 - \kappa \chi_{22}(\omega)}  $, 
$ s_{23}(\omega)=\sqrt{\kappa \gamma_m} \chi_{23}(\omega) $, $ s_{25}(\omega)=\sqrt{\kappa \gamma_d} \chi_{25}(\omega)  $, $ s_{32}(\omega)= \sqrt{\kappa \gamma_m} \chi_{32}(\omega)  $, $  s_{33}(\omega)= 1 -\gamma_m \chi_{33}(\omega) $, $ s_{35}(\omega)= \sqrt{\kappa \gamma_d} \chi_{35}(\omega) $, $ s_{41}(\omega)=\sqrt{\kappa \gamma_m}\chi_{41}(\omega) $, $ s_{44}(\omega)= 1 - \gamma_m \chi_{44}(\omega) $, $ s_{46}(\omega)=\sqrt{\gamma_m \gamma_d} \chi_{46}(\omega) $, $ s_{52}(\omega)= {\sqrt{\kappa \gamma_d} \chi_{52} (\omega)} $$ s_{53}(\omega)= {\sqrt{\gamma_m \gamma_d} \chi_{53}(\omega)} $, $ s_{55}(\omega)= {1 - \gamma_d \chi_{55}(\omega)} $, $ s_{61}(\omega)={\sqrt{\kappa \gamma_d} \chi_{61}(\omega)} $, $ s_{64}(\omega) ={\sqrt{\gamma_m \gamma_d} \chi_{64}(\omega)}  $, and $ s_{66}(\omega)= {1 - \gamma_d \chi_{66}(\omega)} $.

Note that the stability of the system can be determined from the poles of the susceptibility matrix $ \boldsymbol{ { \chi }} (\omega) $ \cite{optomechanicswithtwophonondriving}. In order to keep the system in the stable regime, and for  avoiding mode-splitting the poles should lie in the lower half-plane on the imaginary axis.


\subsection{gain, added noise and amplification }
Using the scattering matrix (\ref{scattering}) together with the symmetrized correlator which is defined as $ \bar S_{AB}^{out}(\omega)=\frac{1}{2} \langle \hat A_{out}(\omega) \hat B_{out}(-\omega) + \hat B_{out}(-\omega) \hat A_{out}(\omega) \rangle $, one can calculate the amplified quadratures of the optical , mechanical, and Bogoliubov modes, respectively, as follows
\begin{subequations} \label{amplification}
\begin{eqnarray}
&& \frac{\bar S_{PP,a}^{out}(\omega)}{\mathcal{G}_a(\omega)}= (\bar n_a^T+\frac{1}{2}) + \bar n_{add,a}^{(amp)}(\omega) , \label{s_pp_a}   \\
&& \frac{\bar S_{XX,m}^{out}(\omega)}{\mathcal{G}_m(\omega)}= (\bar n_m^T+\frac{1}{2}) + \bar n_{add,m}^{(amp)}(\omega), \label{s_xx_m}     \\
&& \frac{\bar S_{XX,d}^{out}(\omega)}{\mathcal{G}_d(\omega)}= (\bar n_d^T+\frac{1}{2}) + \bar n_{add,d}^{(amp)}(\omega) , \label{s_xx_d}
\end{eqnarray}
\end{subequations}
where $ \mathcal{G}_a(\omega)=\vert 1-\kappa \chi_{22}(\omega) \vert^2 $ and  $ \mathcal{G}_{m(d)}(\omega)=\vert 1-\gamma_{m(d)} \chi_{33(55)}(\omega) \vert^2 $ are the frequency-dependent gains. Also, the amplifier-added noises 
are given by
\begin{subequations} \label{added_omega}
\begin{eqnarray}
&& \! \bar n_{add,a}^{(amp)}(\omega)\!=\! \kappa  \mathcal{G}_a^{-1}(\omega) \! \left[(\bar n_m^T \!\! + \! \frac{1}{2}) \gamma_m \vert \chi_{23}(\omega) \vert^2 \!\!  + \! (\bar n_d^T \!\! + \! \frac{1}{2}) \gamma_d \vert \chi_{25}(\omega) \vert^2 \!  \right] , \label{n_add_a_omega} \nonumber   \\
&& \\
&& \! \bar n_{add,m}^{(amp)}(\omega)\! = \! \gamma_m \mathcal{G}_m^{-1}(\omega)\!  \left[ (\bar n_a^T \! \! +\! \frac{1}{2})  \kappa \vert \chi_{32}(\omega) \vert^2 \!\! + \! (\bar n_d^T \!\! + \! \frac{1}{2}) \gamma_d \vert \chi_{35}(\omega) \vert^2 \! \right], \label{n_add_m_omega} \nonumber  \\
&& \\
&&  \! \bar n_{add,d}^{(amp)}(\omega) \! = \! \gamma_d \mathcal{G}_d^{-1}(\omega) \! \left[(\bar n_a^T \!\! + \! \frac{1}{2}) \kappa \vert \chi_{52}(\omega) \vert^2 \! \! +\!  (\bar n_m^T \!\! + \! \frac{1}{2}) \gamma_m\vert \chi_{53}(\omega) \vert^2 \! \right]. \label{n_add_d_omega} \nonumber \\
&& 
\end{eqnarray}
\end{subequations}

On-resonance ($ \omega=0 $), the optical, mechanical and Bogoliubov gains, $ \sqrt{\mathcal{G}_j}\equiv\mathcal{Q}_j/\mathcal{P}_j $ with $ j=a,m,d $, can be simplified as
\begin{subequations} \label{gains}
\begin{eqnarray}
&& \sqrt{\mathcal{G}_a}= \frac{\mathcal{C}_0-(1-\xi_m)+\mathcal{C}_1 \frac{1-\xi_m}{1-\xi_d}}{\mathcal{C}_0+(1-\xi_m)+\mathcal{C}_1 \frac{1-\xi_m}{1-\xi_d}} , \label{g_a}	\\
&& \sqrt{\mathcal{G}_m}= \dfrac{\mathcal{C}_0-(1+\xi_m)-\mathcal{C}_1 \frac{1+\xi_m}{1-\xi_d}}{\mathcal{C}_0+(1-\xi_m)+\mathcal{C}_1 \frac{1-\xi_m}{1-\xi_d}}  , \label{g_m}\\
&& \sqrt{\mathcal{G}_d}=\frac{\mathcal{C}_1-(1+\xi_d)-\mathcal{C}_0 \frac{1+\xi_d}{1-\xi_m} }{\mathcal{C}_1+(1-\xi_d)+\mathcal{C}_0 \frac{1-\xi_d}{1-\xi_m}} . \label{g_d} 
\end{eqnarray}
\end{subequations}
As is evident from Eqs.~(\ref{gains}b) and (\ref{gains}c) under the replacements $ \mathcal{C}_0 \leftrightarrow \mathcal{C}_1 $ and $ \xi_m \leftrightarrow \xi_d $, one has $  \mathcal{G}_m \leftrightarrow \mathcal{G}_d $.
Moreover, on-resonance amplifier-added noises are given by 
\begin{subequations} \label{added_resonance}
	\begin{eqnarray}
	&& \bar n_{add,a}^{(amp)}(0)=(\! 1 \!- \!\frac{1}{\sqrt{\mathcal{G}_a}})^2 \! \left[\! (\frac{1}{2} \! +\bar n_m^T)\frac{\mathcal{C}_0}{(1-\xi_m)^2}\! +\!(\frac{1}{2} \!+\!\bar n_d^T)\frac{\mathcal{C}_1}{(1-\xi_d)^2}  \! \right], \nonumber \\
	&& \\
	&& \bar n_{add,m}^{(amp)}(0)=\frac{4\mathcal{C}_0}{\mathcal{Q}_m^2}\left[ (\bar n_a^T + \frac{1}{2} )+ (\bar n_d^T + \frac{1}{2}) \frac{\mathcal{C}_1}{(1-\xi_d)^2} \right] , \\
	&& \bar n_{add,d}^{(amp)}(0)=\frac{4\mathcal{C}_1}{\mathcal{Q}_d^2}\left[ (\bar n_a^T + \frac{1}{2} )+ (\bar n_m^T + \frac{1}{2}) \frac{\mathcal{C}_0}{(1-\xi_m)^2} \right] .
	\end{eqnarray}
\end{subequations}
From Eqs.(\ref{added_resonance}a)-(\ref{added_resonance}c) one finds that for each of the three modes (optical, mechanical, and Bogoliubov modes) the corresponding added noise includes the contributions originating from the noises of the two other modes. Moreover, as is evident, under the replacements $ \mathcal{C}_0 \leftrightarrow \mathcal{C}_1 $, $ \xi_m \leftrightarrow \xi_d $, and $\bar n_m^T \leftrightarrow \bar n_d^T $  not only the optical added noise $\bar n_{add,a}^{(amp)}(0)$ is invariant, but also $\bar n_{add,m}^{(amp)}(0) \leftrightarrow \bar n_{add,d}^{(amp)}(0) $. This indicates the symmetric roles of the mechanical and Bogoliubov modes in on-resonance optical added noise.

Interestingly, although our system is very complicated compared to a standard DPA but we will show that all added noises can be vanished (analogous to what happens in an ideal DPA) by controlling cooperativities and modulation parameters through the atomic collisions frequency and mechanical spring coefficient. On the other hand, it has been shown \cite{Quantum-Limited-Amplification} that suppression of the mechanical thermal noise contribution to the optical added noise is achievable in the dissipative optomechanical amplification scheme. 

It should be noted that the added noise suppression in the present system is different from that of a simple optomechanical amplifier realized by driving an optomechanical cavity on its blue sideband regime which corresponds to the two-mode squeezing interaction. In fact, in a DPA the mechanical noise is not cooled, and can represent a significant source of added noise for the amplifier. Surprisingly, we will show that, one can achieve simultaneously large gains corresponding to the amplifier together with suppressed added noises by controlling the modulation parameters and cooperativities. In other words, we show that the scattering matrix can be controlled by the system parameters in order to behave as a noiseless quantum-limited phase-sensitive amplifier ($ \bar n_{add,j}^{(amp)} \to 0 $ and $  \mathcal{G}_j \gg 1 $) \cite{Refrevise8} for each subsystem, in which the noiseless amplified subsystem is entirely decoupled from the other subsystems.

\begin{figure} 
\centering
\includegraphics[width=4.23cm]{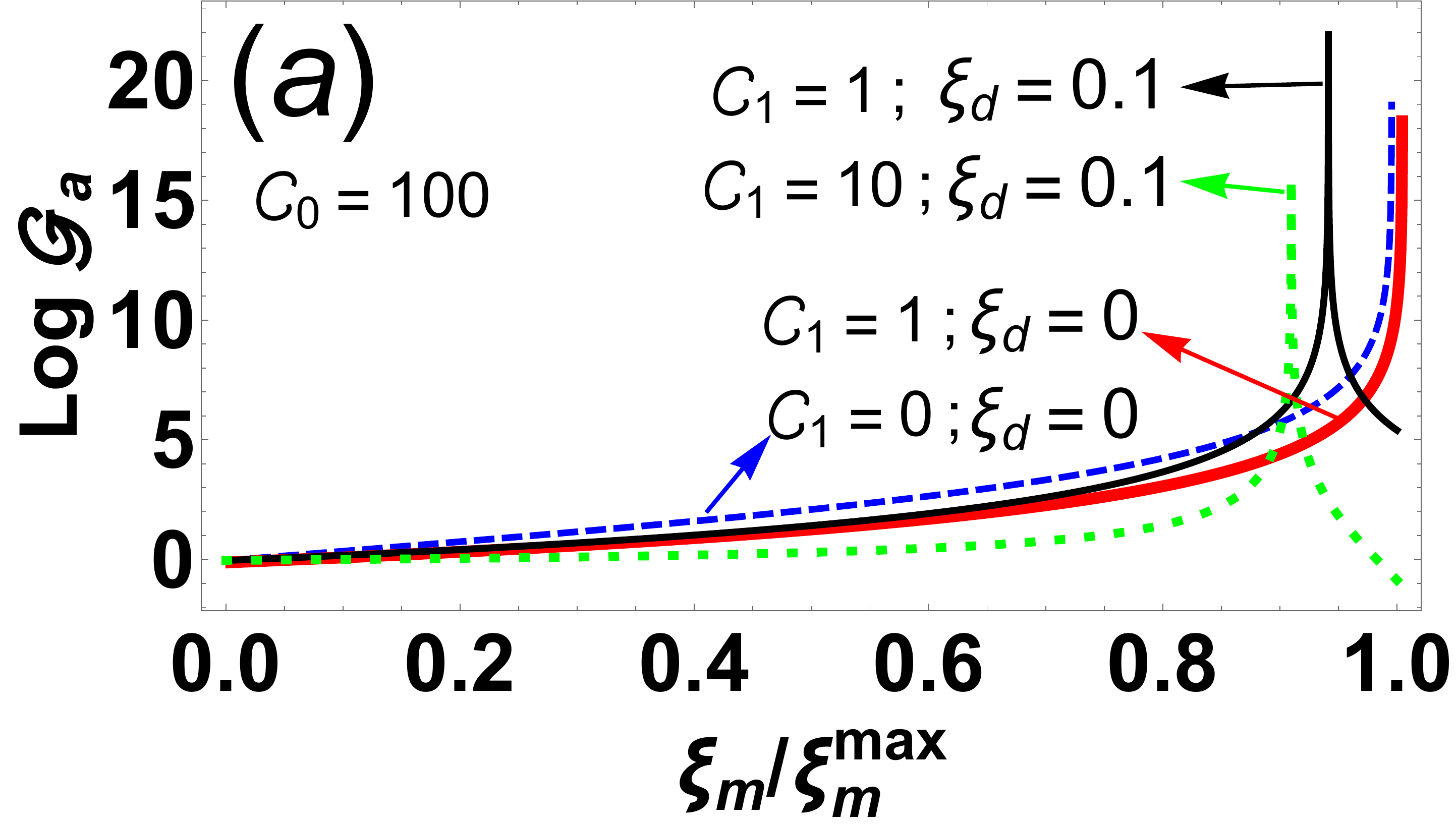} 
\includegraphics[width=4.34cm]{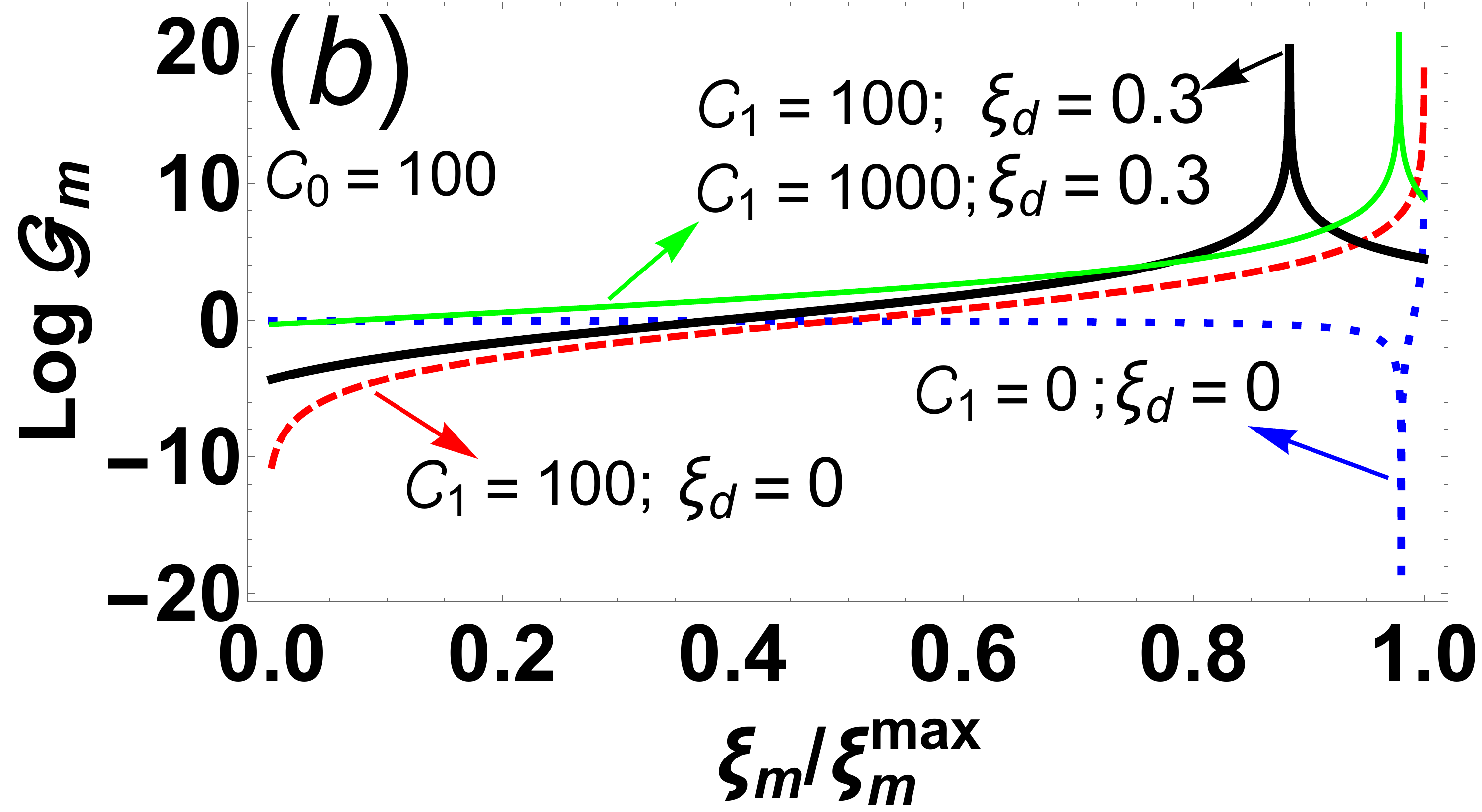} 
\caption{(Color online) On-resonance logarithm of (a) the optical and (b) the mechanical gain vs $ \xi_m/\xi_m^{max} $ for different values of opto-atomic cooperativity $ \mathcal{C}_1 $ and atomic parametric drive $ \xi_d $.  Here, the optomechanical cooperativity is fixed to be $ \mathcal{C}_0=10^2 $.}
\label{fig2}
\end{figure}

Figure~(\ref{fig2}) shows the on-resonance logarithm of optical and mechanical gains versus normalized effective mechanical paramp, $ \xi_m/\xi_m^{max} $, for different values of cooperativity, $ \mathcal{C}_{1} $, and effective atomic paramp, $ \xi_d $. As is seen from Fig.~\ref{fig2}(a), in the absence of the BEC (blue dashed line) or in its presence with no atomic parametric drive (red solid thick line), there is no optical gain except for $\xi_{m}\approx \xi_{m}^{max}$. On the other hand, in the presence of the BEC and with turning on atomic paramp one can obtain a considerable gain for some specified values of the mechanical driving. Especially, in the largely different cooperativities regime, i.e., $ \mathcal{C}_0 \gg \mathcal{C}_1 $ (black thin line), the optical gain shows a high peak at $\xi_{m}\approx 0.9\xi_{m}^{max}$. 

The mechanical gain $ \mathcal{G}_m $ has been shown in Fig.~\ref{fig2}(b). Just like what observed for the optical gain, in the absence of the BEC (blue dotted line) or in its presence with no atomic parametric drive (red dashed line), there is no mechanical gain except for the limit of $ \xi_m \to \xi_m^{max} $. However, by turning on the atomic paramp ($ \xi_d \neq 0 $), for $ \mathcal{C}_1 \geq \mathcal{C}_0 $ (black solid thick and green solid thin lines), one can have much larger mechanical gain in comparison to what obtained in Ref.~\cite{optomechanicswithtwophonondriving} (blue dotted line). Therefore, one of the advantages of the present system in comparison to the one studied in Ref.~\cite{optomechanicswithtwophonondriving}, where the cavity contains no BEC, is the capability of obtaining high optical and mechanical gains by controlling the atomic parameters. 

The appearance of the peaks in the optical and mechanical gains can be explained in terms of the set of Eqs.~(\ref{g_a})-(\ref{g_d}). Here, it should be noted that when the denominators go to zero, i.e., $ \mathcal{P}_j \to 0 $, one can have large gains ($ \mathcal{G}_j \to \infty $) while if the numerators become zero, i.e., $ \mathcal{Q}_j \to 0 $, the gains go to zero. For example for the optical gain, generally $ \xi_m \to \xi_m^{(p,a)} $ and $ \xi_m \to \xi_m^{(Q,a)} $ correspond, respectively, to the large and weak gains in which $ \xi_m^{(p,a)}=1+ \mathcal{C}_0 (1-\xi_d)/ [\mathcal{C}_1+(1-\xi_d) ] $ and $ \xi_m^{(Q,a)}= 1+ \mathcal{C}_0 (1-\xi_d)/ [\mathcal{C}_1-(1-\xi_d )] $ are, respectively, the zeros of the denominator and numerator of the optical gain. As is evident one always has $  \xi_m^{(P,a)} \le \xi_m^{(Q,a)} \le  \xi_m^{max}= 1+\mathcal{C}_m $, and that is why the optical gain in the presence of the atomic paramp is not always a monotonous increasing function of the mechanical paramp $ \xi_m $.

\begin{figure}
	\centering
	\includegraphics[width=4.28cm]{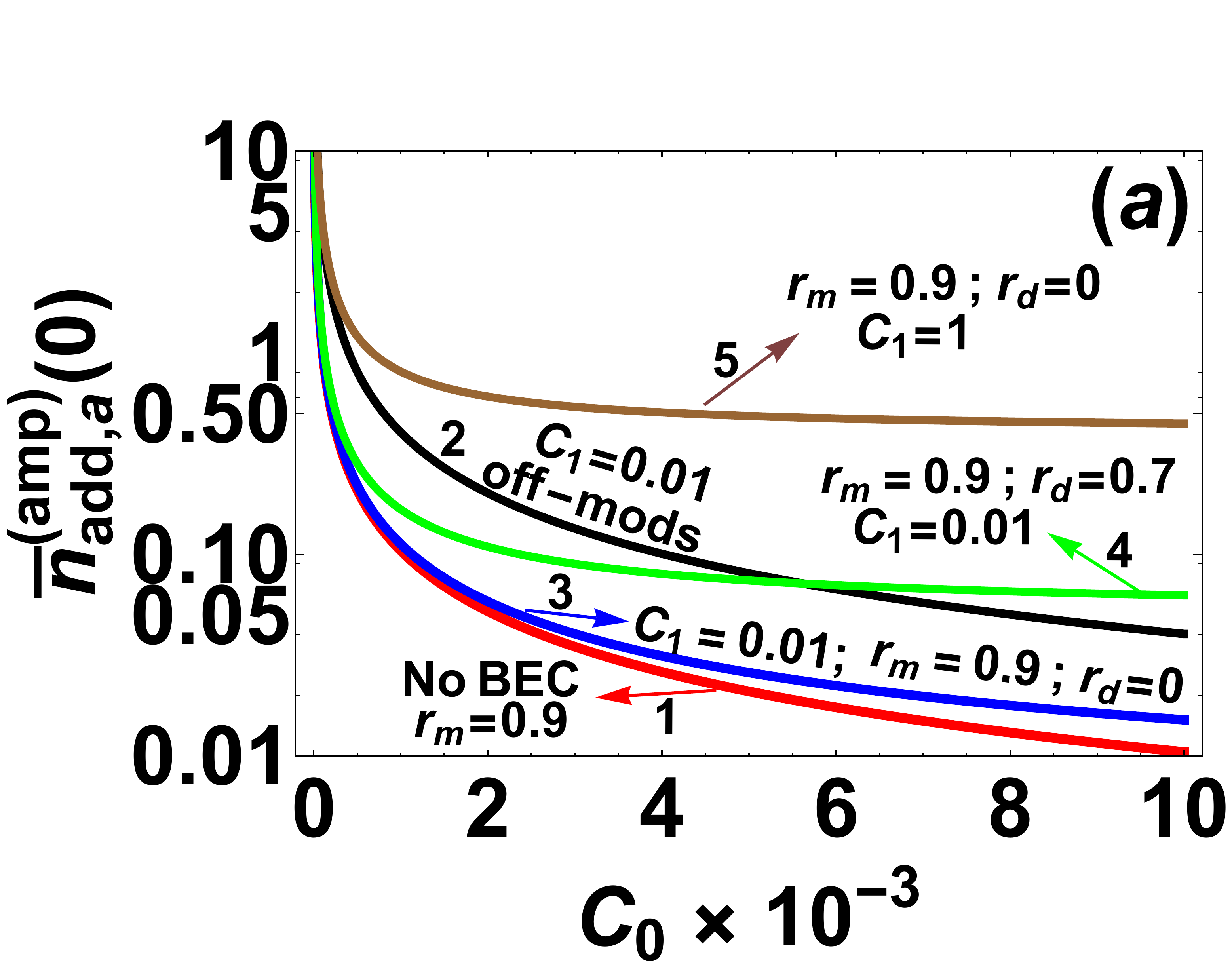} 
	\includegraphics[width=4.29cm]{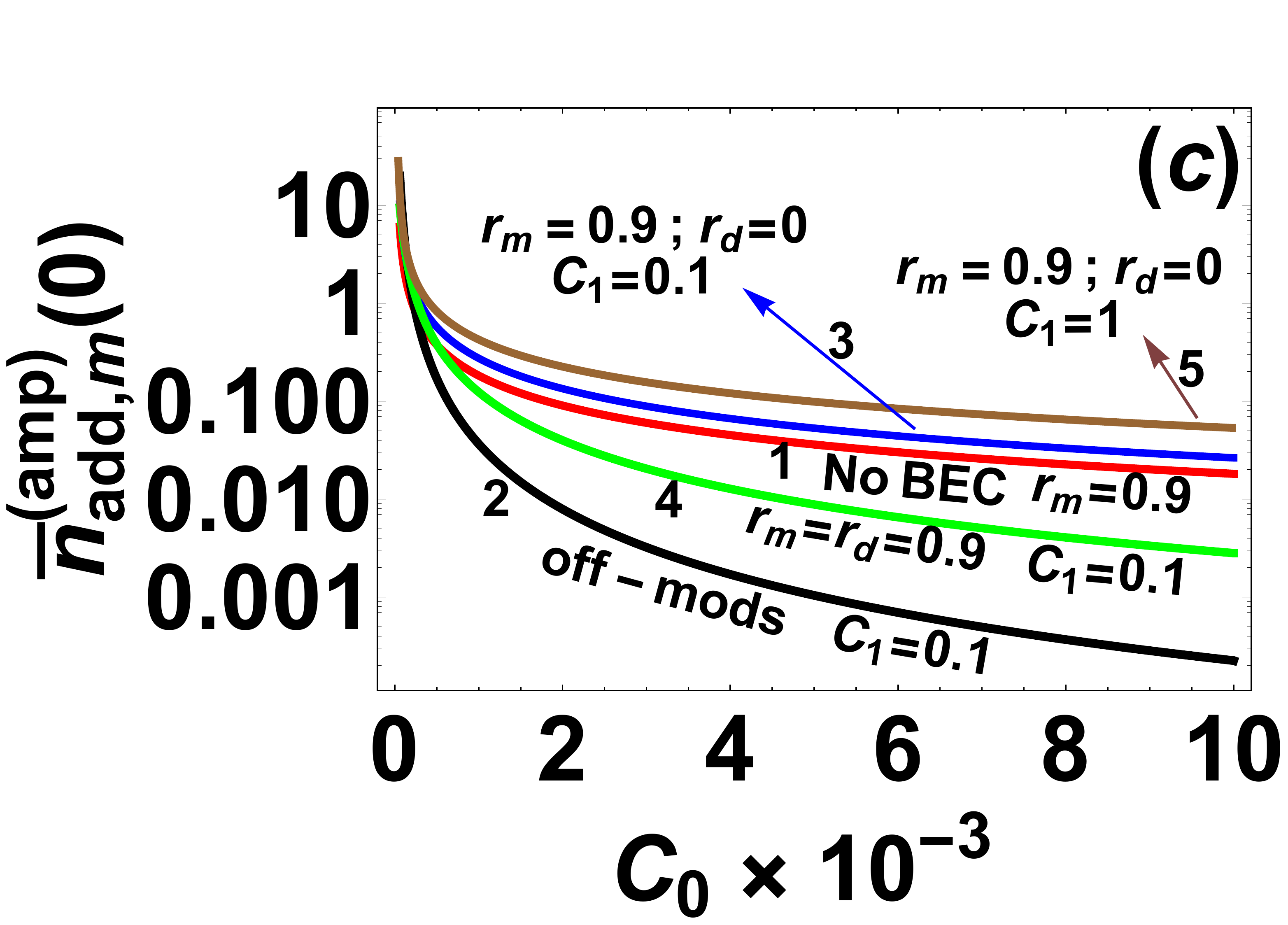} 
	\includegraphics[width=4.28cm]{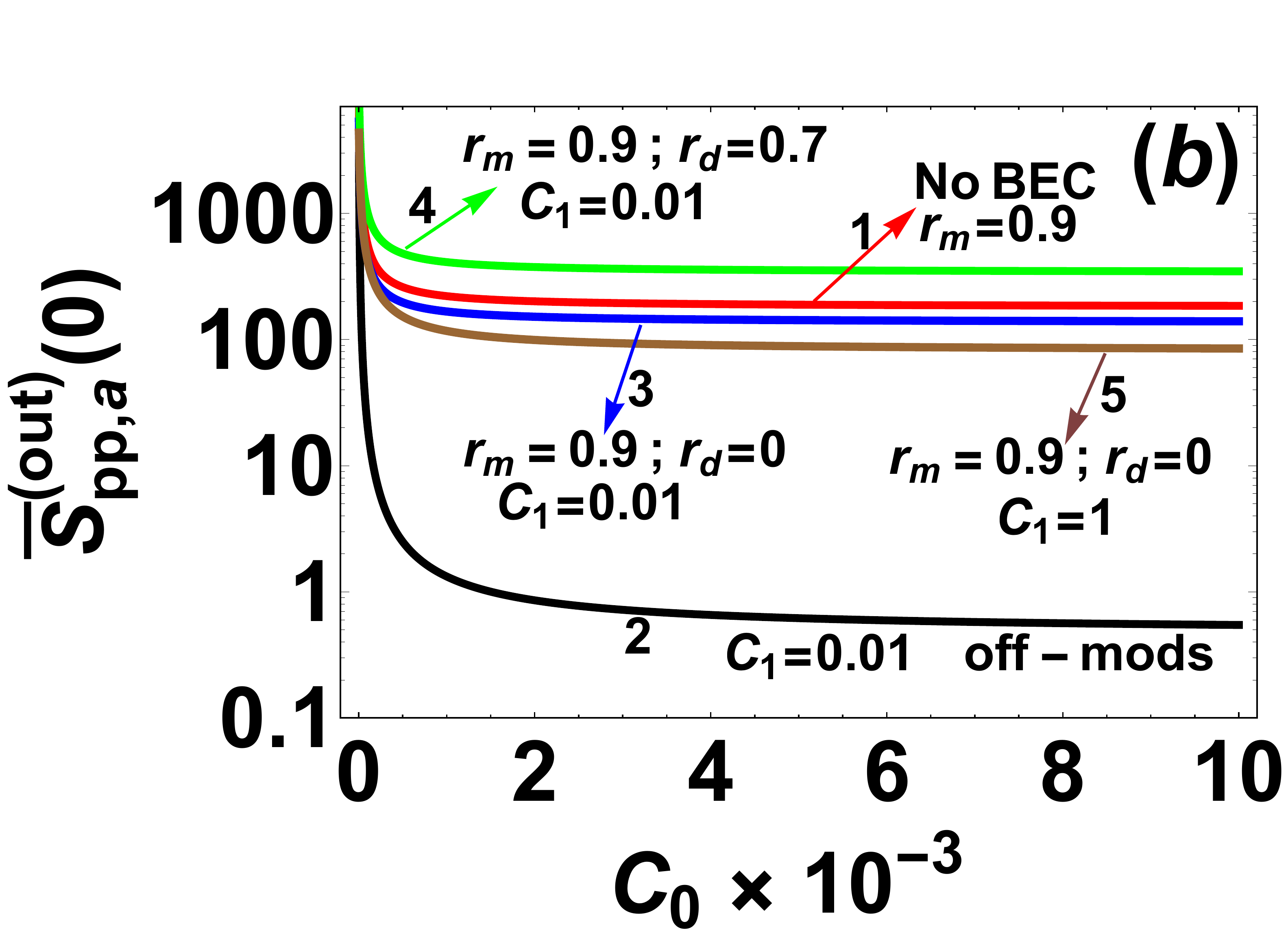}
	\includegraphics[width=4.29cm]{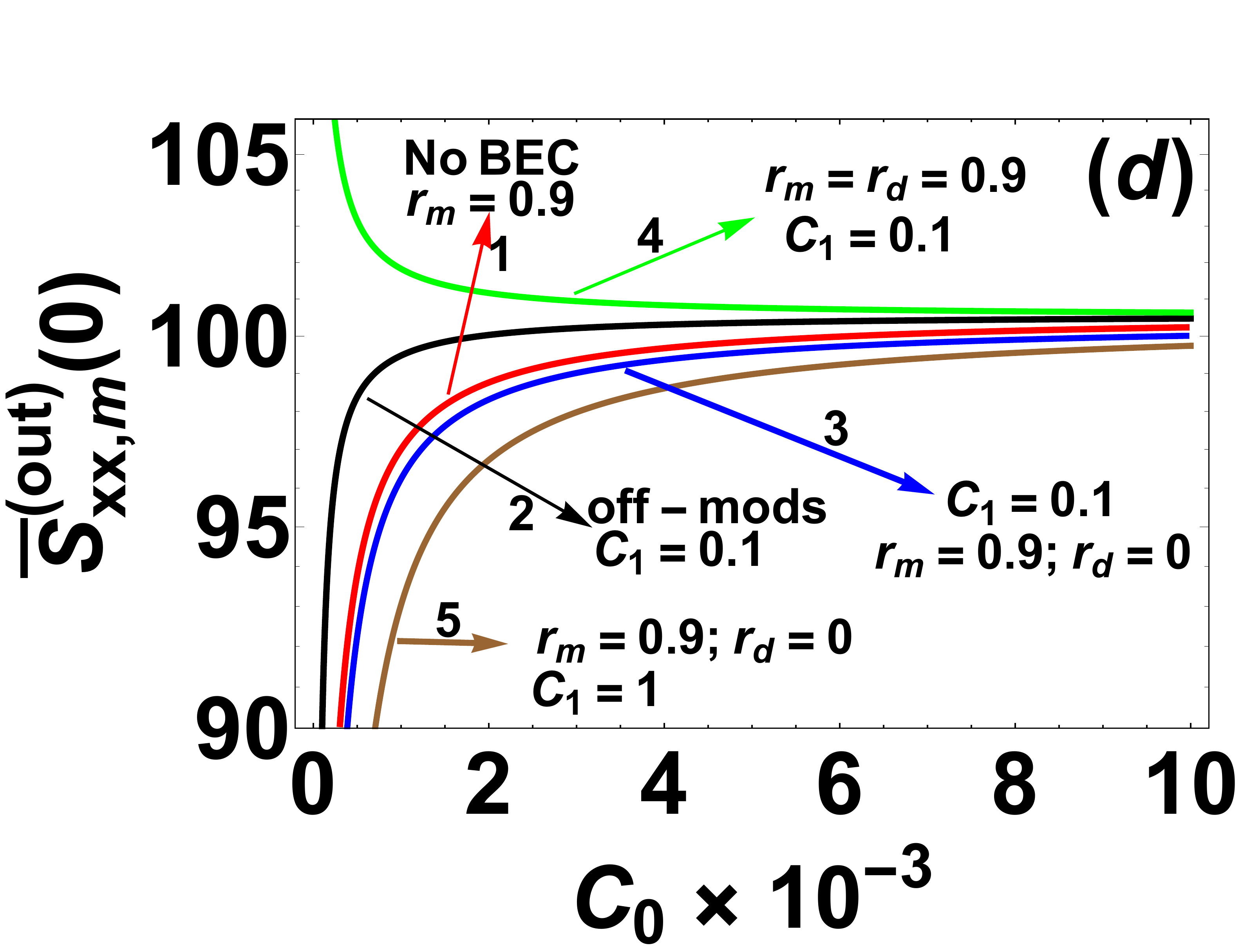} 
	\caption{(Color online) (a) and (c) show the on-resonance optical and mechanical added noises ($ \bar n_{add,a}^{(amp)}(0)$ and $ \bar n_{add,m}^{(amp)}(0) $), respectively, and (b) and (d) show the on-resonance optical and mechanical amplified quadratures spectra, $ \bar S_{PP,a}^{out}(0) $ and $ \bar S_{XX,m}^{out}(0) $, respectively, vs optomechanical cooperativity $ \mathcal{C}_0 $ for different values of $ \mathcal{C}_1 $, $ r_m=\xi_m/\xi_{m}^{max} $ and $ r_d=\xi_d/\xi_{d}^{max} $. Here, $ \bar n_m^T=100 $ and $ \bar n_a^T=\bar n_d^T \simeq 0 $.}
	\label{fig3}
\end{figure}

Figure~(\ref{fig3}) shows the on-resonance added-noises and amplified quadratures spectra versus optomechanical cooperativity, $ \mathcal{C}_0 $, for different values of $ \mathcal{C}_1 $, $ r_m=\xi_m/\xi_{m}^{max} $ and $ r_d=\xi_d/\xi_{d}^{max} $. In each panel the red line indicated by 1 corresponds to the case studied in Ref.~\cite{optomechanicswithtwophonondriving} where there is no BEC in the cavity while the mechanical oscillator is modulated. As is seen, in such a system the optical added noise is very low [Fig.~\ref{fig3}(a)] but due to the mechanical modulation the the P-quadrature of the optical field is effectively amplified [Fig.~\ref{fig3}(b)]. On the other hand, the black line indicated by 2 shows the situation where the cavity contains a BEC but both the mechanical and atomic modulations are turned off (off-mods). Although in this case the optical and mechanical added noises are low [Fig.~\ref{fig3}(a) and \ref{fig3}(c)] but there is no amplification in the P-quadrature of the optical field [Fig.~\ref{fig3}(b)].

More interestingly, for the system equipped with a BEC with both the mechanical and the atomic modulations turned on [the green lines indicated by 4 in Fig.~(\ref{fig3})] the highest optical and mechanical amplification can be achieved [Figs.~\ref{fig3}(b) and \ref{fig3}(d)] while the optical and mechanical added noises are still low enough [Figs.~\ref{fig3}(a) and \ref{fig3}(c)]. This situation is achieved in the largely different cooperativities regime ($ \vert \mathcal{C}_1 - \mathcal{C}_0 \vert \gg 1  $) with weak opto-atom cooperativity ($ \mathcal{C}_1 <1 $). In order to see the destructive effect of large values of opto-atomic cooperativity, one can compare the lines indicated by 3 and 5 which represent similar conditions with different opto-atomic cooperativities $\mathcal{C}_1$ . As is seen, the increase of opto-atomic cooperativity leads to substantial increase in both the optical and the mechanical added noises while it does not much help to the amplifications.

Finally, it should be mentioned that here we have left out the investigation of the atomic amplification and its added noise since, if $\mathcal{C}_1 \leftrightarrow \mathcal{C}_0 $ and $ \xi_d \leftrightarrow \xi_m $, the same results can be obtained for the atomic added noise and atomic amplification.

\subsection{controllable gain-bandwidth}

In the previous subsection leaving out the frequency dependence of the added-noise and quadratures amplification, we studied the details of their on-resonance behaviors [Fig.~(\ref{fig3})]. In this subsection we are going to consider the gain-bandwidth limitation in our system, i.e., the full-width at half maximum (FWHM) of the amplified quadratures (the elements $  \vert s_{jj}[\omega] \vert^2 $ in Eq.~(\ref{scattering}) with $ j=2,3,5 $) specially the optical amplification bandwidth. 
To determine the amplifier bandwidth, we consider the \textit{s}-matrix elements as $ s_{jj}[\omega] \propto 1/D_g[\omega] $ where $ D_g[\omega] $ is a cubic function of $ \omega $, i.e., $ D_g[\omega]=i\omega^3 +a \omega^2 +b \omega +c $ with the coefficients given by 
\begin{eqnarray} \label{bc}
&& 4b \approx \kappa \gamma_m (\mathcal{C}_0+1-\xi_m) +\kappa \gamma_d (\mathcal{C}_1+1-\xi_d) +\bar \gamma^2 \xi_m \xi_d , \nonumber \\
&& c \approx \frac{\kappa \bar \gamma^2}{8} \left[ \mathcal{C}_0(1-\xi_d)+\mathcal{C}_1(1-\xi_m) + 1+\xi_m \xi_d -\xi_m - \xi_d   \right] , \nonumber \\
&& -2a \approx \kappa + \gamma_m(1-\xi_m ) +  \gamma_d(1-\xi_d ) ,
\end{eqnarray}
where $ \bar \gamma=\sqrt{\gamma_m \gamma_d} $. In order to find the FWHM one should solve for the equation $ D_g[\delta \omega]=2D_g[0] $. Under the conditions $ \xi_d=0 $ and $ \xi_m \to \xi_m^{max}=1+\mathcal{C}_0/(1+\mathcal{C}_1) $ one can show straightforwardly that $ c \to 0 $, therefore the gain-bandwidth can be easily calculated as follows
\begin{eqnarray} \label{gainbandwidth1}
&& \Delta \omega_g=-\frac{-2b}{a}= \frac{\kappa \gamma_m (\mathcal{C}_0+1-\xi_m) +\kappa \gamma_d (\mathcal{C}_1+1) }{\kappa + \gamma_m(1-\xi_m ) +  \gamma_d}.
\end{eqnarray}
Here, one can ignore $\gamma_{d}$ because in the normal dissipation regime $ \gamma_d\ll\kappa $. Moreover, in the limit of large optical gain ($ \sqrt{\mathcal{G}_a } \gg 1 $) which corresponds to $ \xi_m \to \xi_m^{max} $, the gain-bandwidth can be written in terms of the optical gain as follows
\begin{eqnarray} \label{gainbandwidth4}
&& \Delta \omega_g \! \! \sqrt{\mathcal{G}_a} \! = \! 2\mathcal{C}_0 \gamma_m \frac{1 \!- \! \frac{\sqrt{\mathcal{G}_a}}{2\mathcal{C}_0} \left[ \mathcal{C}_1 (1-\xi_m) -\dfrac{\gamma_d}{\gamma_m}(1+\mathcal{C}_1) \right]}{1-\frac{\gamma_m }{\kappa}\left[ \mathcal{C}_0+\mathcal{C}_1(1-\xi_m)  \right] }, 
\end{eqnarray}
where $ 2\mathcal{C}_0 \gamma_m=8g^2/\kappa  $. In the weak coupling regime ( $ g,G \ll \kappa $) the gain-bandwidth takes the form
\begin{eqnarray} \label{gainbandwidth-final}
&& \Delta \omega_g \approx \frac{8g^2}{\kappa } \left[\frac{1}{ \sqrt{ \mathcal{G}_a}}  + \frac{\mathcal{C}_1 }{2 \mathcal{C}_0}(\xi_m-1) +\frac{\gamma_d}{\gamma_m}\frac{(1+\mathcal{C}_1)}{2 \mathcal{C}_0}  \right]. 
\end{eqnarray}

As is seen from Eq.~(\ref{gainbandwidth-final}), in the large gain limit and weak atomic modulation, i.e., $ \xi_d \approx 0 $, the gain-bandwidth can be controlled by the ratios $ g/\kappa $, $ \gamma_m/\gamma_d $, the ratio of cooperativities $ \mathcal{C}_{1}/\mathcal{C}_{0} $ and also by the mechanical parametric drive $ \xi_m $. In Eq.~(\ref{gainbandwidth-final}), the second and third terms in the brackets which are due to the presence of the Bogoliubov mode of the BEC, have no counterparts in the bare optomechanical system studied in Ref.~\cite{optomechanicswithtwophonondriving}. The presence of these terms can provide more controllability on the gain-bandwidth of the present hybrid system through the BEC parameters. Furthermore, the presence of the BEC
increases the gain-bandwidth which can be considered as another advantage of our optomechanical amplifier. Note that $\xi_{d}$ can
be controlled by $\omega_{sw}$ and $\epsilon$ while $\xi_{m}$ can be controlled via $\delta k$.

\section{quadratures squeezing spectra \label{sec.squeezing}}
Quadratures squeezing spectra can be investigated by complementary amplified quadratures which are given by
\begin{subequations} \label{squeezing-spectrum}
	\begin{eqnarray}
	&& \frac{\bar S_{XX,a}^{out}(\omega)}{1/2}= (2\bar n_a^T+1) \vert 1-\kappa\chi_{11}(\omega) \vert^2 \nonumber \\
	&& \quad + (2\bar n_m^T+1)  \kappa \gamma_m \vert \chi_{14}(\omega) \vert^2 \! + \! (2\bar n_d^T+1)  \kappa \gamma_d \vert \chi_{16}(\omega) \vert^2 , \label{s_xx_a}   \\
	&& \frac{\bar S_{PP,m}^{out}(\omega)}{1/2}= (2\bar n_m^T+1) \vert 1-\gamma_m\chi_{44}(\omega)\vert^2 \nonumber \\ 
	&& \quad +  (2\bar n_a^T+1)  \kappa \gamma_m \vert \chi_{41}(\omega) \vert^2 \! + \!(2\bar n_d^T+1) \gamma_m \gamma_d \vert \chi_{46}(\omega) \vert^2 ,  \label{s_pp_m}     \\
	&& \frac{\bar S_{PP,d}^{out}(\omega)}{1/2}=  (2\bar n_d^T+1) \vert 1-\gamma_d\chi_{66}(\omega) \vert^2\nonumber \\
	&&\quad  + (2\bar n_a^T+1)  \kappa \gamma_d \vert \chi_{61}(\omega) \vert^2 \! + \! (2\bar n_m^T+1) \gamma_m \gamma_d \vert \chi_{64}(\omega) \vert^2. \label{s_pp_d}
	\end{eqnarray}
\end{subequations}
The on-resonance ($ \omega= 0$) output squeezing spectra are given by 
\begin{subequations} \label{squeezing-on-resonance}
	\begin{eqnarray}
	&&  \!\!\!\!\! \frac{\bar S_{XX,a}^{out}(0)}{1/2}=(1+2\bar n_a^T) \left( 1-\frac{2}{\eta_a} \right)^2 \nonumber \\
	&&  + (\!1+\!2\bar n_m^T) \frac{4\mathcal{C}_0}{(\! 1+\!\xi_m)^2} \frac{1}{\eta_a^2} \!+ \! (1\!+\! 2\bar n_d^T) \frac{4\mathcal{C}_1}{(1+\xi_d)^2} \frac{1}{\eta_a^2}, \label{s_xx_a0}   \\
	&&  \!\!\!\!\! \frac{\bar S_{PP,m}^{out}(0)}{1/2}= \frac{1}{s_m^2} \left[ (1+2\bar n_m^T) \eta_m^2 + 4\mathcal{C}_0 (1+2\bar n_a^T) \right. \nonumber \\
	&&\qquad\qquad\qquad\qquad \left.  + \frac{4\mathcal{C}_0 \mathcal{C}_1}{(1+\xi_d)^2} (1+2\bar n_d^T) \right]  , \label{s_pp_m0}     \\
	&& \!\!\!\!\!  \frac{\bar S_{PP,d}^{out}(0)}{1/2}=  \frac{1}{s_d^2} \left[ (1+2\bar n_d^T) \eta_d^2  + 4\mathcal{C}_1 (1+2\bar n_a^T)  \right. \nonumber \\
	&& \qquad\qquad\qquad\qquad \left. + \frac{4\mathcal{C}_1 \mathcal{C}_1}{(1+\xi_m)^2} (1+2\bar n_d^T) \right],  \label{s_pp_d0}
	\end{eqnarray}
\end{subequations}
with 
\begin{subequations} \label{eta}
	\begin{eqnarray}
	&&  \eta_a=1+\frac{\mathcal{C}_0}{1+\xi_m} +\frac{\mathcal{C}_1}{1+\xi_d}, \label{eta_a}  \\
	&&  \eta_{m(d)}=\mathcal{C}_{0(1)}+\xi_{m(d)}-1+\mathcal{C}_{1(0)} \frac{\xi_{m(d)}-1}{1+\xi_{d(m)}}, \label{eta_m,d}  \\
	&&   s_{m(d)}= \mathcal{C}_{0(1)}+1+\xi_{m(d)} +\mathcal{C}_{1(0)} \frac{1+\xi_{m(d)}}{1+\xi_{d(m)}} . \label{s_m,d}
	\end{eqnarray}
\end{subequations}
Squeezing occurs if the right-hand sides of Eqs.~(\ref{squeezing-spectrum}a)-(\ref{squeezing-spectrum}c) (or Eqs.~(\ref{squeezing-on-resonance}a)-(\ref{squeezing-on-resonance}c)) become smaller than unity. In addition to the degree of squeezing which is characterized by the quadrature spectrum, the purity or impurity of squeezing is another important feature. The impurity of output squeezing spectrum can be characterized by the effective thermal occupancy \cite{optomechanicswithtwophonondriving} which is given by ($j=a,m,d$)
\begin{eqnarray}  \label{Impurity frequency}
&&\!\!\!\! (\bar n_{{\rm eff}, j}^{\rm out} [\omega] \! + \! \frac{1}{2})^2 \! =\bar S_{XX,j}^{out}[\omega] \bar S_{PP,j}^{out}[\omega]- \bar S_{XP,j}^{out}[\omega] \bar S_{PX,j}^{out}[\omega] ,
\end{eqnarray}
where the cross correlations are given by 
\begin{subequations} \label{cross correlations}
	\begin{eqnarray}
	&& \bar S_{XP,a}^{out}[\omega]=\bar S_{PX,a}^{out}[\omega]= -\frac{1}{2} {\rm Im} \left[ (1-\kappa \chi_{11}(\omega)) (1-\kappa \chi_{22}^\ast(\omega))  \right. \nonumber \\
	&& \qquad\qquad\quad \left. +\kappa\gamma_m\chi_{23}(\omega) \chi_{14}^\ast(\omega)+ \kappa\gamma_d\chi_{25}(\omega) \chi_{16}^\ast(\omega) \right], \\
	&& \bar S_{XP,m}^{out}[\omega]=\bar S_{PX,m}^{out}[\omega]= -\frac{1}{2} {\rm Im} \left[ (1-\gamma_m \chi_{33}(\omega)) (1-\gamma_m \chi_{44}^\ast(\omega))  \right. \nonumber \\
	&& \qquad\qquad\quad \left.+\gamma_m\kappa\chi_{41}(\omega)\chi_{32}^\ast(\omega)+ \gamma_m \gamma_d\chi_{35}(\omega) \chi_{46}^\ast(\omega) \right], \\
	&& \bar S_{XP,d}^{out}[\omega]=\bar S_{PX,d}^{out}[\omega]= -\frac{1}{2} {\rm Im} \left[ (1-\gamma_d \chi_{55}(\omega)) (1-\gamma_d \chi_{66}^\ast(\omega))  \right. \nonumber \\
	&& \qquad\qquad\quad \left.+\gamma_d\kappa\chi_{61}(\omega)\chi_{52}^\ast(\omega)+ \gamma_d \gamma_m\chi_{53}(\omega) \chi_{64}^\ast(\omega) \right].
	\end{eqnarray}
\end{subequations}
It can be easily shown that in on-resonance ($ \omega=0 $) the cross correlations become zero, i.e., $ \bar S_{xp,j}^{out}[0]=0 $, which is because of the RWA. It should be noted that the impurity $ \bar n_{{\rm eff},j}^{\rm out} [\omega] $ for any pure state of the output mode will be zero, and for a thermal state will be equal to the actual thermal occupancy of that mode \cite{optomechanicswithtwophonondriving}. In the following, we will show how our system has the potential to achieve simultaneously perfect squeezing with high purity.

\subsection{quadrature squeezing of the output cavity field}

Let us show how we can overcome the decoherence effects in order to achieve perfect squeezing for the cavity output field. As can be seen from Eq.~(\ref{s_xx_a0}), in the limit $\eta_a\to 2$  the contribution of the optical noise in the on-resonance output optical squeezing spectrum vanishes and thus  
\begin{eqnarray} \label{cavity squeezing 1}
&& \frac{\bar S_{XX,a}^{out}(0)}{1/2} \to (\!1+\!2\bar n_m^T) \frac{\mathcal{C}_0}{(\! 1+\!\xi_m)^2} \!+ \! (1\!+\! 2\bar n_d^T) \frac{\mathcal{C}_1}{(1+\xi_d)^2}. 
\end{eqnarray}
It is easy to verify that in the absence of atomic modulation $(\xi_{d}=0)$ together with strong mechanical driving such that $ \xi_m \to \xi_{m,sq}^{a,max}=[(1+\mathcal{C}_1)\xi_m^{max}-2]/(1-\mathcal{C}_1) $, the limit $\eta_a\to 2$  is achievable. Under such conditions, we get
\begin{eqnarray} \label{cavity squeezing final}
&& \frac{\bar S_{xx,a}^{out}(0)}{1/2} \to (1-\mathcal{C}_1)^2 \frac{(\!1+\!2\bar n_m^T) }{\mathcal{C}_0}+ \! \mathcal{C}_1 (1\!+\! 2\bar n_d^T)  . 
\end{eqnarray}

As is seen from Eq.~(\ref{cavity squeezing final}), in the regime of $ \mathcal{C}_1 \ll 1 $ and $ \mathcal{C}_0 \gg 1 $ (i.e., the regime of largely different cooperativities), and when $\bar n_d^T$ is negligibly small \cite{BrennBECexp,RitterBECexp}, the squeezing of the on-resonance cavity output quadrature $\hat X_{out}(0)$ below zero-point is set by the ratio $\bar n_m^T/\mathcal{C}_0$ . This means that significant degree of squeezing can be achieved by ground state cooling of the MO. However, in the limit $\bar n_m^T\ll \mathcal{C}_0$, which corresponds to the regime of strong optomechanical coupling between the MO and the intracavity field, i.e., when $g^2\gg\frac{\kappa\gamma_{m}}{4}\frac{k_B T_m}{\hbar\omega_{m}}$, one can achieve strong output field squeezing which is robust against thermal noises. It should be mentioned that interchanging the mechanical and the Bogoliubov modes ($ \xi_m \leftrightarrow \xi_d, \mathcal{C}_0 \leftrightarrow \mathcal{C}_1, \bar n_m^T\leftrightarrow\bar n_d^T $) leaves the results unchanged.

\begin{figure}
	\centering
	\includegraphics[width=4.25cm]{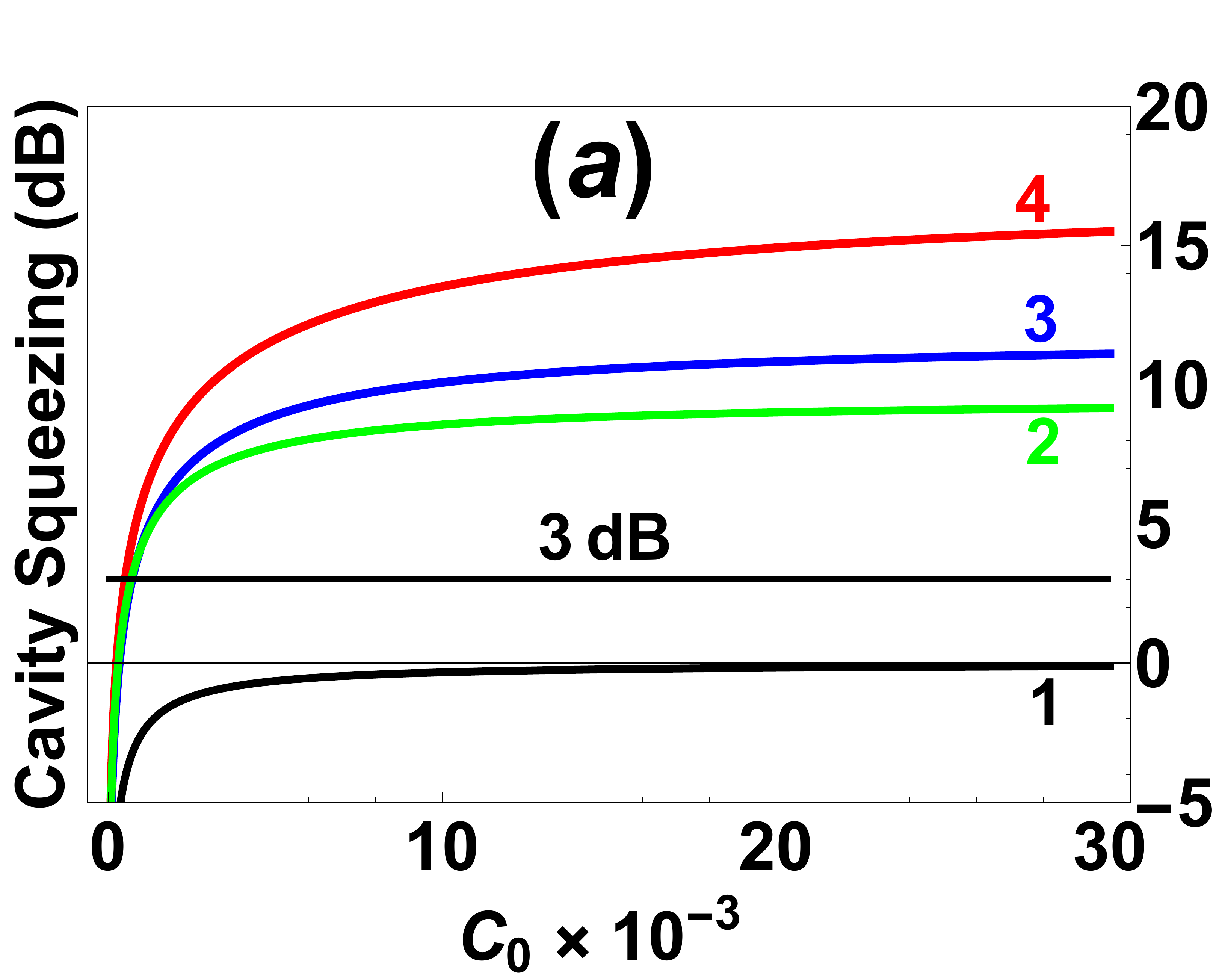} 
	\includegraphics[width=4.32cm]{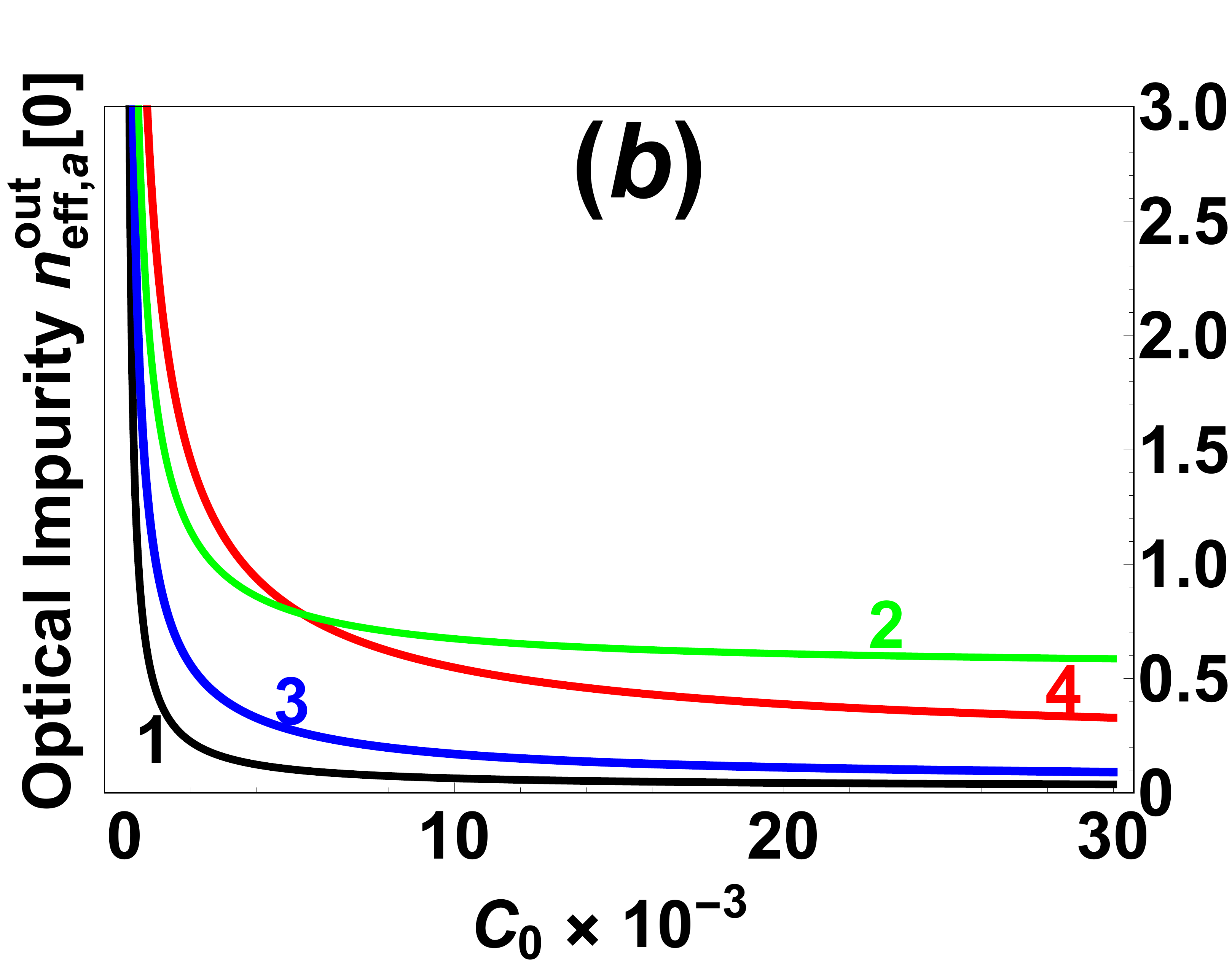} 
	\caption{(Color online)
		(a) Squeezing and (b) purity of the output cavity field on-resonance vs optomechanical cooperativity $ \mathcal{C}_0 $ in the absence of atomic modulation ($\xi_d=0 $) for different values of $\xi_{m}$ and $ \mathcal{C}_1 $. The lines marked by 1, 2, 3, and 4 correspond, respectively, to the cases ($ \xi_{m} = 0, \mathcal{C}_1=10^{-2} $), ($ \xi_{m} = 0.6\xi_{m}^{max}, \mathcal{C}_1=10^{-1} $), ($ \xi_{m} = 0.6\xi_{m}^{max}, \mathcal{C}_1=10^{-2} $), and ($ \xi_{m} = 0.8\xi_{m}^{max},\mathcal{C}_1=10^{-2} $).  Here, we have set $ \bar n_m^T=100 $ and $ \bar n_a^T=\bar n_d^T=0 $.
	 }
	\label{fig4}
\end{figure}

In Figs.~(\ref{fig4}a) and (\ref{fig4}b) we have plotted, respectively, the degree of squeezing of the output cavity field on-resonance ($-10\log_{10}2\bar S_{XX,a}^{out}(0)$ , in dB unit) and its purity ($\bar n^{out}_{{\rm eff},a}(0)$) versus the optomechanical cooperativity $\mathcal{C}_0$ in the absence of atomic modulation ($\xi_{d}=0$) for different values of $\xi_{m}$ and $\mathcal{C}_1$ ($\mathcal{C}_1\ll 1$). As is evident, in the largely different cooperativities regime ($ \vert \mathcal{C}_0-\mathcal{C}_1 \vert \gg 1 $), one can beat the 3dB limit ($50\%$ squeezing) through the increase of mechanical modulation amplitude ($\xi_{m}$) without losing the purity of squeezing. Also, by increasing the difference between cooperativities both the degree of squeezing and its purity increase. However, in spite of the cavity squeezing enhancement, its purity decreases a little bit with increasing the mechanical modulation.

In fact, there is a competition between the maximum achievable squeezing and its purity which can be controlled by the difference between the cooperativities and the effective amplitude of the mechanical modulation $ \xi_m $. 
This trade-off can be interpreted as follows: while heating the MO and BEC via the coherent modulation they are cooled by the intracavity light field through the radiation pressure in the red-detuned regime. These phononic fluctuations are transferred into the P-quadrature of the light field via the optomechanical coupling and therefore that trade-off originating from cooling and heating of the BEC and MO is induced into the cavity output-field squeezing and its purity. 

It should be noted that the maximum achievable squeezing can be even more than 16 dB which never occurs in the standard OMSs. On the other hand, as can be seen from Fig.~(\ref{fig4}), the generated squeezing of the output cavity field and its purity are robust to the mechanical thermal noises. Moreover, the high degree of optical squeezing in our system can be attributed to the \textit{induced} frequency-dependent squeezing [see Eq.~(\ref{lambda_a_omega})] because of the cavity field interaction with BEC and MO in the presence of the external modulations of the MO and atomic collisions.

\subsection{quadrature squeezing of the mechanical phonons}
Here, we show that in the present hybrid optomechanical system one can achieve perfect quadrature squeezing of the phononic mode of the MO or of the BEC with mediocre purity via controlling the atomic or mechanical modulation as well as cooperativities such that both conditions $\eta_m\to 0$ and $s_m\gg 1$ are satisfied.

To this end, we consider two special cases: (i) no atomic collisions modulation ($ \xi_d=0 $) and (ii) no mechanical frequency modulation ($ \xi_m=0 $). In the first case, if  $ \xi_m \to 2-\xi_m^{max} $ then according to Eqs.~(\ref{eta_m,d}) and (\ref{s_m,d}), $\eta_m\to 0$ and $ s_m \to 2(1+\mathcal{C}_1) $. Therefore in the limit of large opto-atomic cooperativity ($\mathcal{C}_1\gg 1$) one has $s_m\gg 1$ , and consequently the squeezing of the on-resonance P-quadrature of the MO below zero-point [Eq.~(\ref{s_pp_m0})] reads as
\begin{eqnarray} \label{mechanical squeezing 2}
&& \frac{\bar S_{PP,m}^{out}(0)}{1/2} \to \frac{\mathcal{C}_0}{\mathcal{C}_1^2} (1+2\bar n_a^T)+ \frac{\mathcal{C}_0}{\mathcal{C}_1} (1+2\bar n_d^T)  ,
\end{eqnarray}
which shows that for $\mathcal{C}_0\ll\mathcal{C}_1$ together with $\bar n_a^T=\bar n_d^T\approx 0$, one can achieve high degree of mechanical P-quadrature squeezing much above the standard quantum limit of 3dB limit. On the other hand, in the second case, i.e., $\xi_m=0$, if $ \xi_d \to [(1+\mathcal{C}_0)\xi_d^{max}-2\mathcal{C}_0]/(\mathcal{C}_0-1)=(\mathcal{C}_1+1-\mathcal{C}_0)/(\mathcal{C}_0-1)  $ (with $ \mathcal{C}_0 >1 $ and $\mathcal{C}_1>\mathcal{C}_0-1$) then $\eta_m\to 0$ and $s_m\to 2\mathcal{C}_0$. In this case, we get
\begin{eqnarray} \label{mechanical squeezing 3}
&& \frac{\bar S_{PP,m}^{out}(0)}{1/2} \to \frac{(1+2\bar n_a^T)}{\mathcal{C}_0} + \frac{\mathcal{C}_0-1}{\mathcal{C}_1} (1+2\bar n_d^T)  ,
\end{eqnarray}
which clearly shows that the squeezing of the P-quadrature of the MO can be significantly enhanced by requiring $\mathcal{C}_1\gg\mathcal{C}_0\gg 1$.
Note that in both cases the perfect mechanical squeezing can be achieved in the largely different cooperativities regime ($ \vert \mathcal{C}_0-\mathcal{C}_1 \vert \gg 1 $). It should be mentioned that because of the symmetry between the two phononic modes, i.e., mechanical and Bogoliubov modes, the same results can be obtained for the Bogoliubov-type phonon squeezing with the replacements $ \mathcal{C}_0 \leftrightarrow \mathcal{C}_1 $ and $ \xi_m  \leftrightarrow  \xi_d$.

\begin{figure}
	\centering
	\includegraphics[width=4.25cm]{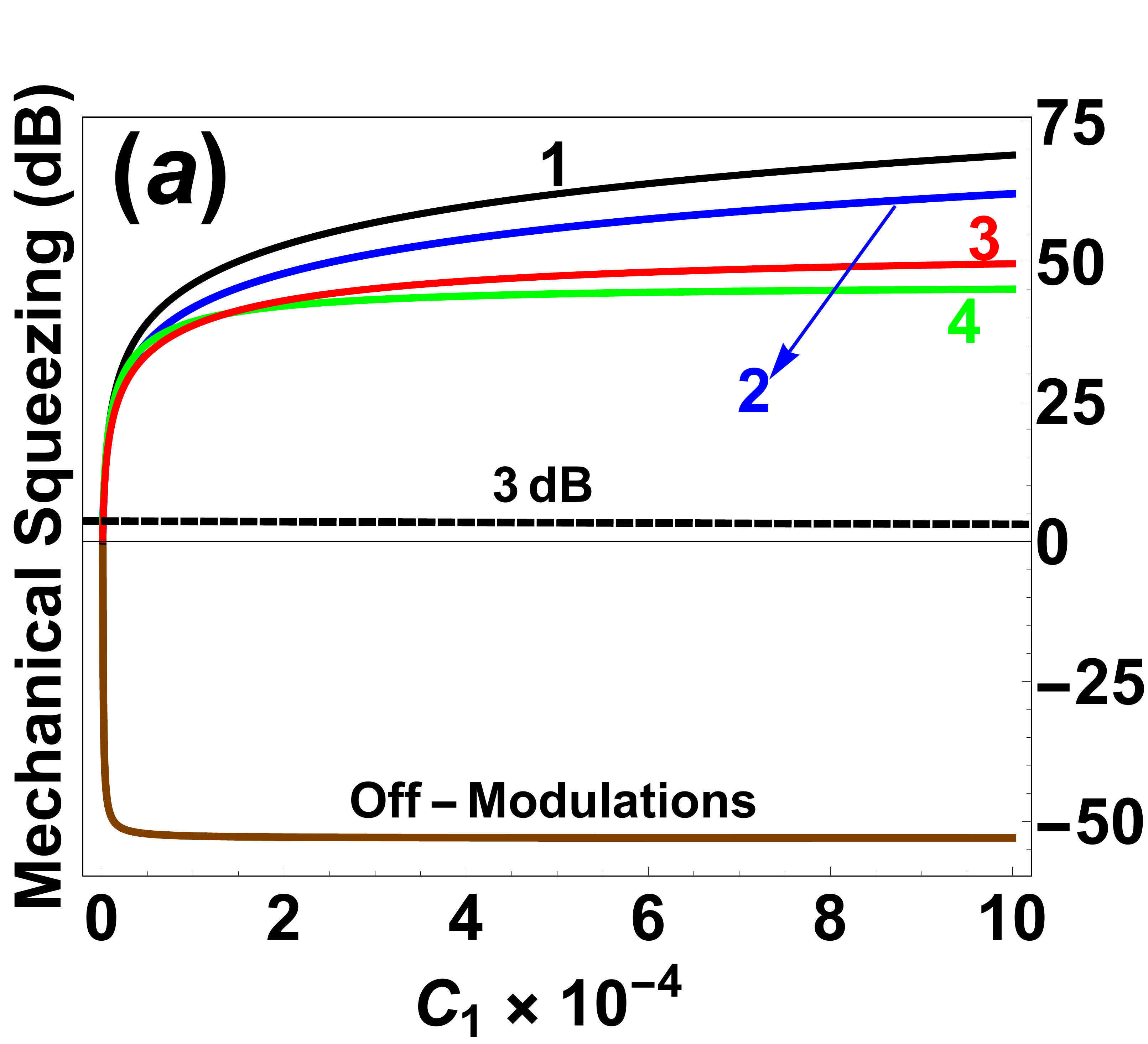} 
	\includegraphics[width=4.32cm]{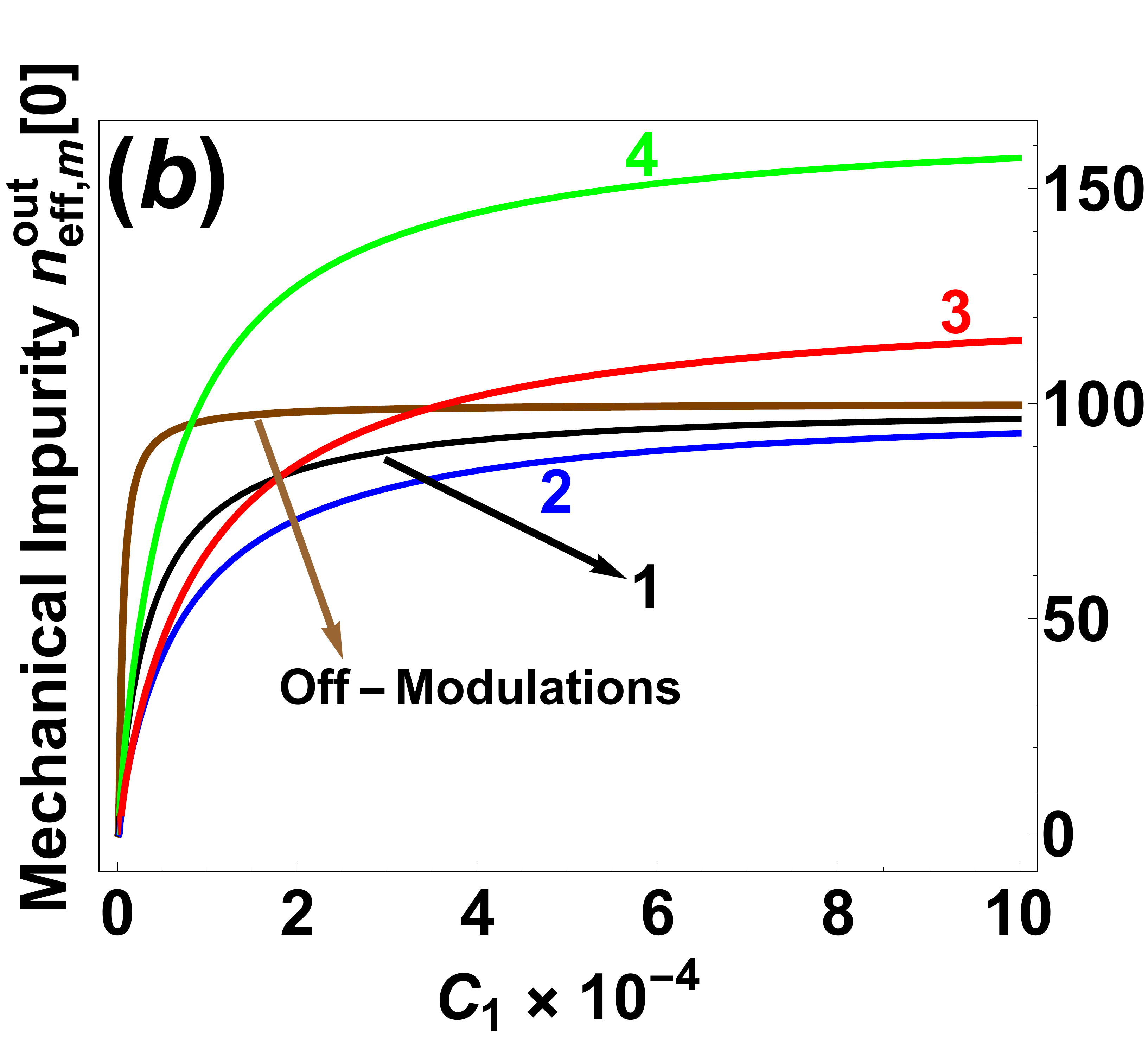} 
	\caption{(Color online) (a) On-resonance mechanical squeezing, $ -10\log2\bar S_{PP,m}^{out}(0)$, and (b) on-resonance mechanical impurity of state, $ \bar n_{{\rm eff},m}^{\rm out} [0] $, vs opto-atomic cooperativity $ \mathcal{C}_1 $. The brown lines correspond to the off-modulations ($ \xi_{m}=\xi_d=0 $) for $ \mathcal{C}_0=100 $. The black and blue lines, respectively, indicated by 1 and 2 correspond to the case (i) in the text, i.e., off-atomic modulation $ \xi_d=0 $, for $ \mathcal{C}_0=100$ and $ \mathcal{C}_0=200$, respectively. The red and green lines, respectively, indicated by 3 and 4 correspond to the case (ii) in the text, i.e., off-mechanical modulation $ \xi_m=0 $, for $ \mathcal{C}_0=200 $ and $ \mathcal{C}_0=100 $, respectively. Here, we have set $ \bar n_m^T=100 $ and $ \bar n_a^T=\bar n_d^T = 0 $.}
	\label{fig5}
\end{figure}

Figure (\ref{fig5}) shows the on-resonsnce mechanical $P$-quadrature squeezing and its impurity versus opto-atomic cooperativity $\mathcal{C}_1$ in the red-detuned and weak coupling regimes. As is seen, there is no mechanical squeezing in the absence of modulations [brown line indicated by off-modulations] while the squeezing is generated as soon as the modulations are turned on and increases rapidly by increasing the opto-atomic cooperativity $ \mathcal{C}_{1} $ so that it surpasses the 3 dB limit in the largely different cooperativities regime. On the other hand, case (i), i.e., off-atomic modulation with large mechanical parametric drive leads to much more degree of squeezing than case (ii), i.e., off-mechanical modulation with large atomic parametric drive (compare lines 1 and 2 with lines 3 and 4 in Fig.~\ref{fig5}(a).

Moreover, by increasing the difference between the cooprativities the impurity increases [Fig.~\ref{fig5}(b)]. In the regime of large parametric drive, since $\bar n_{{\rm eff},m}^{\rm out} [0] \sim \bar n_m^T  $, one can conclude that the state of the MO is a thermal squeezed state which is in agreement with explanations given in Refs.~\cite{aliDCE3,JLiNJP,JLiPRA} based on the collective mode in which two phononic subsystems evolve to a collective Bogoliubov mode as a thermal-squeezed and entangled state. In other words, the generated controllable strong mechanical squeezing reflects to overcome of the noise amplification in the $ X $-quadratures of the MO and BEC to the noise suppression in the $P $-quadratures of the MO and BEC.

It should be pointed out that in spite of the standard ponderomotive squeezing which is maximized in the unresolved sideband regime or bad-cavity limit ($ \kappa \gg \omega_m $), our system operates in the weak couplings, resolved-sideband (good-cavity limit) regimes  ($ g,G < \kappa\ll\omega_m \approx \omega_d $) where very high degree of squeezing (up to 75 dB) is achievable in the red detuned and largely different cooperativities regimes. Nevertheless, the standard ponderomotive squeezing mechanism can lead to higher purities than those obtained in our proposed scheme. Moreover, in the present scheme the squeezed quadrature angle can be controlled through the phases of mechanical and atomic modulations ($\phi_m$ and $\phi_{d}$), while in the ponderomotive approach the squeezing angle is dependent on frequency. Furthermore, the squeezing bandwidth in our system can be controlled by cooperativities, $ g, G/\kappa $, and $ \gamma_m/\gamma_d $. In comparison to Ref.~\cite{optomechanicswithtwophonondriving}, the additional coupling to the atomic BEC in the presence of atomic collisions frequency modulation lead to increase of the gain-bandwidth [see Eq.~(\ref{gainbandwidth-final})] which can be considered as an advantage of our proposed system.

Finally, we should emphasize that the present scheme is experimentally feasible because it is compatible with the experimental data applied in Refs.\cite{BrennBECexp, RitterBECexp} as has been shown in our previous theoretical papers \cite{dalafi2,dalafi6,aliDCE3}. For this purpose, one can consider $ N=10^5 $ Rb atoms inside an optical cavity of length $ L=178 \mu$m with bare frequency $ \omega_{c}=2.41494\times 10^{15} $Hz corresponding to a wavelength of $ \lambda=780 $nm and damping rate $ \kappa=2\pi\times 1.3 $MHz. The atomic $ D_{2} $ transition corresponding to the atomic transition frequency $ \omega_{a}=2.41419\times 10^{15} $Hz couples to the mentioned mode of the cavity. The atom-field coupling strength $ g_{0}=2\pi\times 14.1 $MHz and the recoil frequency of the atoms is $ \omega_{R}=23.7 $KHz. The movable end mirror can be assumed to have a mass of $ m=10^{-9}  $g and damping rate of $ \gamma_{m}=2\pi\times 100 $Hz which oscillates with frequency $ \omega_{m}=10^5 $Hz. Besides, the coherent modulation of the mechanical spring coefficient of the MO and also the time modulation of the \textit{s}-wave scattering frequency of atom-atom interaction of the BEC can be realized experimentally as have been reported, respectively, in \cite{pontinmodulation} and \cite{JaskulaBECmodulationexp}.


\section{concluding remarks \label{sec.summary} }

In summary, we have theoretically proposed and investigated a feasible experimental scheme to generate controllable strong quadrature squeezing and noiseless phase-sensitive amplification in a hybrid optomechanical cavity containing an interacting cigar-shaped atomic BEC in the red-detuned and weak coupling regimes through coherent modulations of the atomic collisions frequency of the BEC and of the mechanical spring coefficient of the MO.

We have shown that by controlling the system parameters such as the amplitudes of modulations and cooperativities, the $ X- $quadrature of the cavity mode and the $ P- $quadrature of the MO/BEC can be strongly squeezed (up to 16 dB and 75 dB for the output cavity field and the MO, respectively) in the largely different cooperativities regime with large amplitude of modulations which never occurs in the red-detuned regime of standard cavity optomechanics. Moreover, the output cavity field is squeezed with considerable amount of purity while the squeezing of the MO/BEC is a thermal-like squeezed state. Furthermore, we have shown that the generated squeezings in the subsystems are robust against the thermal noises. 

On the other hand, by analyzing the complementary quadratures we have shown that the $ P- $quadrature of the cavity mode and the $ X- $quadrature of the MO/BEC can be strongly amplified with simultaneous suppression of added-noises, i.e., noiseless phase-sensitive quantum amplifications. What we should mention as another advantageous feature of the present system is the controllability of the gain-bandwidth and the possibility of increasing it through the modulation of the \textit{s}-wave scattering frequency of the BEC which is experimentally controllable by the transverse trapping frequency.

\appendix

\section{\label{App.A} definition of the elements of the susceptibility matrix of Eq.~(\ref{susceptibility tensor}) }
In Eq.~(\ref{susceptibility tensor}), the matrix elements $ \chi_{ij}(\omega) $ are given by 
\begin{widetext}
	\begin{eqnarray}
	&&\!\!\!\!\!\!\!\!\!\!\! \chi_{11}(\omega)= \left[ \chi_0^{-1}(\omega)+g^2 \chi_{+m}+G^2 \chi_{+d}(\omega)    \right]^{-1};  
	\qquad\qquad\qquad\quad \chi_{14}(\omega)=-g \left[ \chi_0^{-1}(\omega) \chi_{+m}^{-1}(\omega) \! +g^2 \! \! +G^2 \! \chi_{+d}(\omega)  \chi_{+m}^{-1}(\omega)   \right]^{-1} ,\nonumber\\
	&&\!\!\!\!\!\!\!\!\!\!\! \chi_{16}(\omega)=G \left[ \chi_0^{-1}(\omega) \chi_{+d}^{-1}(\omega) \! +G^2 \! \! +g^2 \! \chi_{+m}(\omega)  \chi_{+d}^{-1}(\omega)   \right]^{-1};    
	\qquad\quad \chi_{22}(\omega)=  \left[ \chi_0^{-1}(\omega)+g^2 \chi_{-m}+G^2 \chi_{-d}(\omega)    \right]^{-1} ,\nonumber\\
	&&\!\!\!\!\!\!\!\!\!\!\!\chi_{23}(\omega)=g \left[ \chi_0^{-1}(\omega) \chi_{-m}^{-1}(\omega) \! +g^2 \! \! +G^2 \! \chi_{-d}(\omega)  \chi_{-m}^{-1}(\omega)   \right]^{-1};   
	\qquad\quad \chi_{25}(\omega)=-G \left[ \chi_0^{-1}(\omega) \chi_{-d}^{-1}(\omega) \! +G^2 \! \! +g^2 \! \chi_{-m}(\omega)  \chi_{-d}^{-1}(\omega)   \right]^{-1}, \nonumber\\
	&&\!\!\!\!\!\!\!\!\!\!\! \chi_{32}(\omega)=-g \left[  \chi_0^{-1}(\omega) \chi_{-m}^{-1}(\omega) + g^2 + G^2 \chi_{-d}(\omega) \chi_{-m}^{-1}(\omega)   \right]^{-1}; 
	\quad\quad \chi_{33}(\omega)=\frac{ 1+G^2 \chi_0(\omega) \chi_{-d}(\omega) }{\chi_{-m}^{-1}(\omega) + g^2 \chi_0(\omega) + G^2 \chi_0(\omega) \chi_{-d}(\omega) \chi_{-m}^{-1}(\omega) }, \nonumber\\
	&&\!\!\!\!\!\!\!\!\!\!\! \chi_{35}(\omega)=\frac{gG}{ \chi_0^{-1}(\omega) \chi_{-m}^{-1}(\omega)  \chi_{-d}^{-1}(\omega) + G^2 \chi_{-m}^{-1}(\omega) + g^2 \chi_{-d}^{-1}(\omega)  } ;
	\qquad \chi_{41}(\omega)=g \left[  \chi_0^{-1}(\omega) \chi_{+m}^{-1}(\omega) + g^2 + G^2 \chi_{+d}(\omega) \chi_{+m}^{-1}(\omega)   \right]^{-1}  , \nonumber\\
	&& \!\!\!\!\!\!\!\!\!\!\! \chi_{44}(\omega)=\frac{ 1+G^2 \chi_0(\omega) \chi_{+d}(\omega) }{\chi_{+m}^{-1}(\omega) + g^2 \chi_0(\omega) + G^2 \chi_0(\omega) \chi_{+d}(\omega) \chi_{+m}^{-1}(\omega) };
	\qquad \chi_{46}(\omega)=\frac{gG}{ \chi_0^{-1}(\omega) \chi_{+m}^{-1}(\omega)  \chi_{+d}^{-1}(\omega) + G^2 \chi_{+m}^{-1}(\omega) + g^2 \chi_{+d}^{-1}(\omega)  } , \nonumber\\
	&& \!\!\!\!\!\!\!\!\!\!\! \chi_{52}(\omega)=G \left[  \chi_0^{-1}(\omega) \chi_{-d}^{-1}(\omega) + G^2 + g^2 \chi_{-m}(\omega) \chi_{-d}^{-1}(\omega)   \right]^{-1}   ;
	\qquad \chi_{53}(\omega)=\frac{gG}{ \chi_0^{-1}(\omega) \chi_{-m}^{-1}(\omega)  \chi_{-d}^{-1}(\omega) + G^2 \chi_{-m}^{-1}(\omega) + g^2 \chi_{-d}^{-1}(\omega)  } , \nonumber\\
	&& \!\!\!\!\!\!\!\!\!\!\! \chi_{55}(\omega)=\frac{ 1+g^2 \chi_0(\omega) \chi_{-m}(\omega) }{\chi_{-d}^{-1}(\omega) + G^2 \chi_0(\omega) + g^2 \chi_0(\omega) \chi_{-m}(\omega) \chi_{-d}^{-1}(\omega) };
	\qquad \chi_{61}(\omega)=-G \left[  \chi_0^{-1}(\omega) \chi_{+d}^{-1}(\omega) + G^2 + g^2 \chi_{+m}(\omega) \chi_{+d}^{-1}(\omega)   \right]^{-1}  , \nonumber\\
	&& \!\!\!\!\!\!\!\!\!\!\! \chi_{64}(\omega)=\frac{gG}{ \chi_0^{-1}(\omega) \chi_{+m}^{-1}(\omega)  \chi_{+d}^{-1}(\omega) + G^2 \chi_{+m}^{-1}(\omega) + g^2 \chi_{+d}^{-1}(\omega)  };
	\qquad \chi_{66}(\omega)=\frac{ 1+g^2 \chi_0(\omega) \chi_{+m}(\omega) }{\chi_{+d}^{-1}(\omega) + G^2 \chi_0(\omega) + g^2 \chi_0(\omega) \chi_{+m}(\omega) \chi_{+d}^{-1}(\omega) }, \nonumber \\
	\end{eqnarray}
\end{widetext}
with $ \chi_0^{-1}(\omega)= \kappa/2 - i\omega  $, $ \chi_{\pm m}^{-1}(\omega)= \gamma_m/2 \pm \lambda_m - i \omega  $ and $ \chi_{\pm d}^{-1}(\omega)= \gamma_d/2 \pm \lambda_d - i \omega $. It is clear that $ \chi_0(-\omega)= \chi_0(\omega)^\ast $ and $ \chi_{\pm m,\pm d}(-\omega)= \chi_{\pm m,\pm d}(\omega)^\ast $, so $ \chi_{ij}(-\omega)= \chi_{ij}(\omega)^\ast  $.

\end{document}